\documentclass[aps,PRA,twocolumn,superscriptaddress,10pt,twoside]{revtex4-2} 

\usepackage{amsmath,amssymb}
\usepackage{mathrsfs}
\usepackage{epstopdf}
\usepackage{graphicx}
\usepackage{bm,color,bbm}
\usepackage{natbib}
\usepackage{multirow}
\usepackage[hidelinks]{hyperref}
\usepackage{caption}
\usepackage[table, x11names]{xcolor}
\definecolor{aqua}{rgb}{0.7,1,1}
\definecolor{yellowish}{rgb}{1,1,0.7}
\definecolor{greenish}{rgb}{0.5,1,0.4}
\usepackage{subcaption}
\usepackage{fancyhdr}
\usepackage{multirow}
\usepackage{cancel}
\newcommand{\nn}{\nonumber}

\newcommand{\appropto}{\mathrel{\vcenter{\offinterlineskip\halign{\hfil$##$\cr\propto\cr\noalign{\kern2pt}\sim\cr\noalign{\kern-2pt}}}}}

\begin{document}
\fancyhead{} 
\fancyhead[LE,RO]{\ifnum\value{page}<2\relax\else\thepage\fi}

\title{Principles for Optimizing Quantum Transduction \\ in Piezo-Optomechanical Systems}

\author{James Schneeloch}
\email{james.schneeloch.1@afrl.af.mil}
\affiliation{Air Force Research Laboratory, Information Directorate, Rome, New York, 13441, USA}

\author{Erin Sheridan}
\affiliation{Air Force Research Laboratory, Information Directorate, Rome, New York, 13441, USA}

\author{A. Matthew Smith}
\affiliation{Air Force Research Laboratory, Information Directorate, Rome, New York, 13441, USA}

\author{Christopher C. Tison}
\affiliation{Air Force Research Laboratory, Information Directorate, Rome, New York, 13441, USA}

\author{Daniel L. Campbell}
\affiliation{Air Force Research Laboratory, Information Directorate, Rome, New York, 13441, USA}

\author{Matthew D. LaHaye}
\affiliation{Air Force Research Laboratory, Information Directorate, Rome, New York, 13441, USA}

\author{Michael L. Fanto}
\affiliation{Air Force Research Laboratory, Information Directorate, Rome, New York, 13441, USA}

\author{Paul M. Alsing}
\affiliation{Air Force Research Laboratory, Information Directorate, Rome, New York, 13441, USA}

\date{\today}

\begin{abstract}
Two-way microwave-optical quantum transduction is essential to connecting distant superconducting qubits via optical fiber, and to enable quantum networking at a large scale. In Bl\'esin, Tian, Bhave, and Kippenberg's article, ``Quantum coherent microwave-optical transduction using high overtone bulk acoustic resonances" (Phys. Rev. A, 104, 052601 (2021)), they lay out a two-way quantum transducer converting between microwave photons and telecom-band photons by way of an intermediary GHz-band phonon mode utilizing piezoelectric and optomechanical interactions respectively (and are the first to work out the quantum piezoelectric coupling). In this work, we examine both the piezoelectric, and optomechanical interactions from first principles, and together with the evanescent coupling between optical modes, discuss what parameters matter most in optimizing this kind of quantum transducer. For its additional utility, we have also compiled a table of relevant properties of optical materials that may be used as elements in transducers.
\end{abstract}


\maketitle
\thispagestyle{fancy}

\newpage

\section{Introduction}\label{SectionIntroduction}
Quantum transduction (i.e., where quantum information is transferred from one medium to another) is both a critical and fundamental process needed to connect different species of qubits in a quantum network. This makes more diverse heterogeneous quantum networks possible, and also allows us to leverage the complementary advantages \cite{wang2022quantum,delaney2022superconducting,krastanov2021optically,han2021microwave,lambert2020coherent,awschalom2021development,lauk2020perspectives,tian2015optoelectromechanical,xiang2013hybrid,kurizki2015quantum} of different species of qubits. In particular, superconducting qubits, which function at microwave frequencies, enable rapid and deterministic gate operations, while photon-based qubits allow one to send quantum information long distances over optical fiber at ambient temperatures with relatively low loss (e.g., $0.18$dB/km for SMF28-Ultra 200 optical fiber at $1550$nm \cite{CorningFiberSheet}) compared to sending microwave signals over coaxial cables (of the order $10^{2}$ to $10^{3}$dB/km). However, the (microwave/GHz scale) frequencies of light that interact best with superconducting qubits are \emph{five} orders of magnitude below the (telecom) frequencies at which light propagates best in fiber optics. Moreover, the amount of random thermally generated microwave photons that happen to be in a propagating microwave mode compared to a single-photon microwave signal is large enough that resolving a transduced microwave signal from noise requires a cryogenic operating environment \footnote{As a figure of merit, the expected number of $3.285$ GHz microwave photons in a single mode at $298$K, assuming Boltzmann statistics, would be about $1.89\times 10^{3}$, compared to about $0.26$ at $100$mK.}. Unlike classical networking, quantum signals cannot be simply amplified for long-distance transmission due to the no-cloning theorem \cite{wootters1982single}. Linear amplification of quantum signals (possibly written on individual photons) cannot be carried out without adding noise \cite{RevModPhys.82.1155, PhysRevA.86.032106}. This noisy amplification to compensate for loss eventually reaches an irreparable threshold where specific and distinct quantum states (e.g., non-Gaussian states) cannot be reliably transmitted over long distances. In order to connect distant superconducting qubits --- as is needed to establish a network of superconducting quantum computers --- it is necessary to \emph{coherently} convert microwave-photons to telecom photons and back again with as little loss/noise as possible.

In \cite{blesin2021quantum,blesin2023bidirectional}, the authors present and then demonstrate a model of a system that can transduce photons from the microwave frequency band to the telecom frequency band using a high-quality-factor acoustic mode to mediate a pair of coupling interactions. First, light in the form of microwave photons is converted into sound in the form of GHz-scale phonons via the piezoelectric interaction. Then, these phonons are up-converted into telecom-band photons using an optomechanical interaction known as the strain-optical or photoelastic effect, where the extra energy for the upconversion is provided by annihilating photons from a driving telecom-band laser field. By utilizing resonant enhancement of both coupling interactions (to couple the microwave and acoustic modes on resonance at the same GHz- scale frequency, and to couple the acoustic mode to the splitting between the telecom-band resonances in coupled micro-ring resonators), they are able to increase the interaction strength, and in so doing, improve transduction efficiency compared to previous approaches.

In this work, we begin by discussing the essential elements underlying the piezoelectric and optomechanical interactions in order to determine what factors are most easily adjusted to maximize the efficiency of transduction. First, we give a brief overview of the system in question, showing how the transduction efficiency is obtained as a function of the coupling interactions. Following this, we review these coupling constants (derived ab initio in the appendices), and explore what material parameters and device design principles can be used to maximize both the transduction efficiency as well as the bandwidth over which efficient transduction can be accomplished, while minimizing noise.

\section{The Piezo-optomechanical transducer}
\subsection{The physical system}
The piezo-optomechanical transducer in Refs.~\cite{blesin2021quantum,blesin2023bidirectional} consists of multiple elements. 

\begin{figure}[t]
\centerline{\includegraphics[width=0.95\columnwidth]{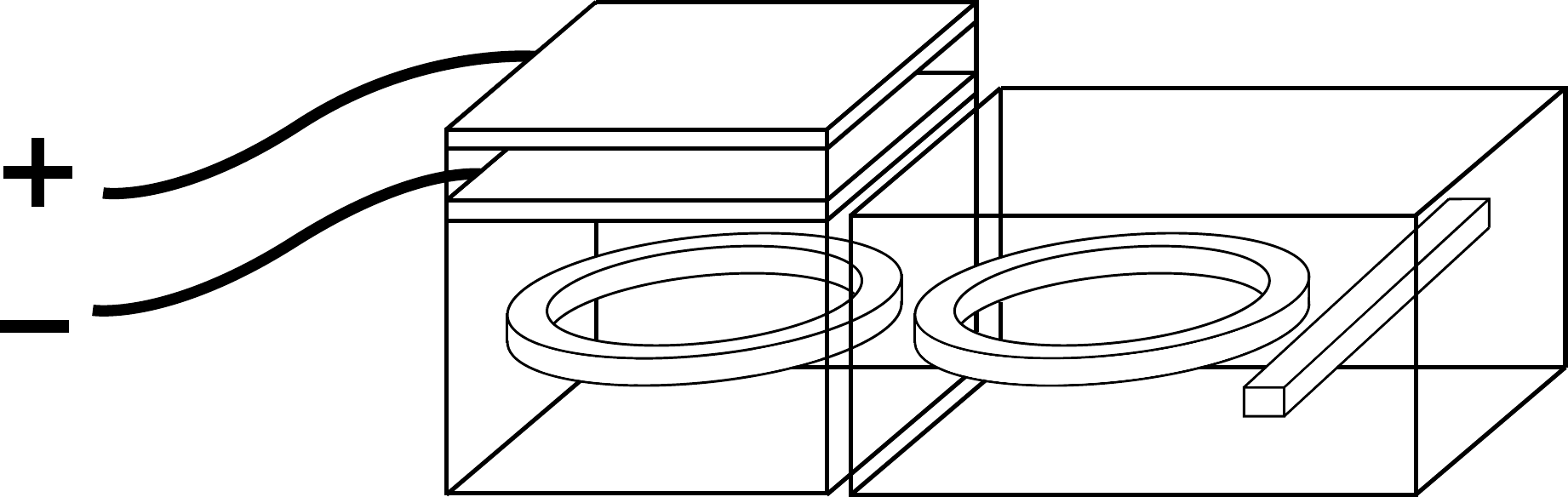}}
\caption{Basic diagram showing the elements of the piezo-optomechanical transducer in \cite{blesin2021quantum} (components not drawn to scale). From left to right: A microwave transmission line is  coupled to opposite ends of a piezoelectric medium bonded to a substrate inside of which a micro-ring resonator (MRR) is embedded. Outside the medium is a second MRR evanescently coupled to the first, which is in turn coupled to a linear bus optical waveguide.}\label{CrudeDiagram}
\end{figure}

Two electrode plates are attached to either end of a flat piezoelectric medium, connecting them to the positive and negative exit terminals of a microwave transmission line. With this, microwave signals traveling to the electrodes can be converted into acoustic vibrations of the piezoelectric medium. See Fig.~\ref{CrudeDiagram} for a simplified diagram of the system.

Bonded to the piezoelectric medium (on the other side of the electrode), is a cladding layer in which a micro ring resonator (MRR) is embedded. This MRR is made of a material sensitive to optomechanical (i.e., photoelastic or acousto-optic) interactions while also serving to confine and guide light within it. In this way, mechanical vibrations in the piezoelectric medium, propagate through the cladding layer and cause the MRR to vibrate as well, which through the optomechanical interaction, modulates the light propagating inside it. Together, we may call this module of the transducer as the bulk acoustic resonator.

The micro-ring resonator inside the cladding layer  (what we will call the cladding resonator, or sometimes, the ``first" resonator) is next to a (nominally) identical \footnote{For microring resonators that cannot truly be \emph{identical} at the time of manufacture, we consider them nominally identical if their independent resonance spectra overlap beyond any Rayleigh criterion to distinguish two peaks.} micro-ring resonator just outside the substrate (what we will call the external, waveguide, or ``second" resonator), with a gap separating the two allowing for evanescent optical coupling between the two resonators.

At the end stage of the transducer, the external resonator embedded in its own cladding layer is evanescently coupled (optically) to an optical bus waveguide. Laser light at a telecom frequency is supplied to the bus waveguide, which is then coupled into the transducer. The optomechanical interaction takes this pump light together with mechanical vibrations propagating through the transducer to produce transduced light at a frequency equal to the sum of the (optical) pump and microwave frequencies. The transduced light then exits through the bus waveguide to be separated from the pump and can then be used in a fiber-optic-connected quantum network.

\subsection{Theoretical model of transduction efficiency}
Quantum-mechanically, we can describe the transduction process (in the single-mode approximation) with the following Hamiltonian:
\begin{align}\label{transducerhamiltonian}
\hat{H} &= \hbar \omega_{m} \hat{b}^{\dagger}\hat{b} + \hbar \omega_{c} \hat{c}^{\dagger}\hat{c} +\hbar \omega_{1} \hat{a}_{1}^{\dagger}\hat{a}_{1} + \hbar \omega_{2} \hat{a}_{2}^{\dagger}\hat{a}_{2}\nn\\
&\qquad +\hbar g_{EM}(\!\hat{c}\!+\!\hat{c}^{\dagger}\!)(\!\hat{b}\!+\!\hat{b}^{\dagger}\!)  - \hbar \bar{G}\hat{a}_{1}^{\dagger}\hat{a}_{1}\!\! \left(\!\hat{b} \!+\! \hat{b}^{\dagger}\!\right)\nn\\
&\qquad - \hbar J\!\! \left(\!\hat{a}_{1}^{\dagger}\hat{a}_{2} \!+\! \hat{a}_{2}^{\dagger}\hat{a}_{1}\!\right)
\end{align}
with definitions of symbols given in Table \ref{table1}:
\begin{table}[!h]
\begin{center}
\begin{tabular}{|c|c|}
\hline
\multicolumn{2}{|c|}{Transduction Hamiltonian symbols}\\
\hline
\hline
    $\omega_{m}$ & Mechanical resonance frequency  \\
    \hline
    $\hat{b}$ & mechanical (phonon) annihilation operator \\
    \hline
    $\omega_{c}$ & microwave photon frequency \\
    \hline
    $\hat{c}$ & microwave photon annihilation operator \\
    \hline
    $\omega_{1}$ & substrate MRR resonance frequency \\
    \hline
    $\hat{a}_{1}$ & substrate MRR photon annihilation operator \\
    \hline
     $\omega_{2}$ & external MRR resonance frequency \\
    \hline
    $\hat{a}_{2}$ & external MRR photon annihilation operator \\
    \hline
    $g_{EM}$ & electromechanical (piezoelectric) coupling rate \\
    \hline
    $\bar{G}$ & single-photon optomechanical coupling rate \\
    \hline
    $J$ & optical (evanescent) coupling rate \\
\hline
\end{tabular}
\caption{List of symbols in transduction Hamiltonian and their definitions. See Table~\ref{table2} for nominal parameters.}\label{table1}
\end{center}
\end{table}

\begin{figure}[t]
\centerline{\includegraphics[width=0.95\columnwidth]{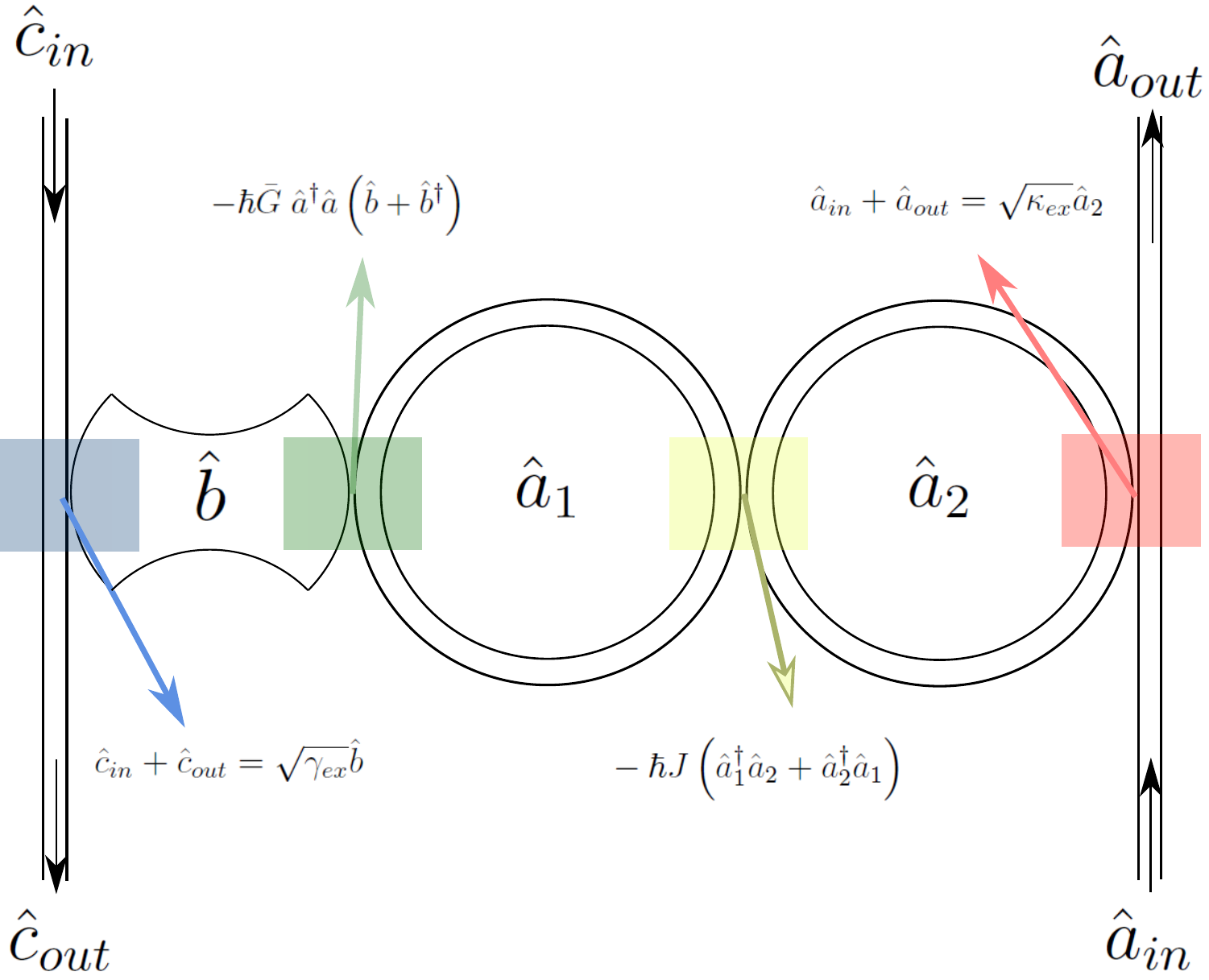}}
\caption{Diagram showing the mode couplings in a quasi-schematic of a piezo-opto-mechanical transducer. The microwave mode $\hat{c}_{\text{\text{in}}}$ couples to the mechanical mode $\hat{b}$ piezoelectrically. The mechanical mode $\hat{b}$ couples to the optical mode $\hat{a}_{1}$ in the cladding MRR optomechanically. The external MRR optical mode $\hat{a}_{2}$ is coupled both to $\hat{a}_{1}$ and $\hat{a}_{\text{out}}$ through evanescent optical coupling interactions.}\label{ModeSchematic}
\end{figure}

Here, the Hamiltonian is broken down into terms as follows. The first four terms are equal to the free field energies of the acoustic field, the microwave field, and the two optical fields in their respective resonators. The fifth term describes the piezoelectric interaction, which couples the microwave and mechanical modes. The sixth term describes the optomechanical interaction, coupling mechanical vibrations with the optical field in the substrate MRR. The final term describes the evanescent coupling interaction between the two MRRs. In Fig.~\ref{ModeSchematic}, we give a simple diagram showing where the different coupling interactions appear within the system.

Not captured in the transduction Hamiltonian are the input, output, external microwave, and optical bosonic mode fields. As these fields are of bosons, we can use standard input-output relations  \cite{WallsMilburn2008,gardiner2004quantum} to phenomenologically introduce loss and noise terms into the Heisenberg-Langevin equations for the fields within the transducer. In addition, we also assume that optical backscattering is negligible, so that we need not consider reverse-propagating optical modes, which would double the complexity of the resulting dynamics for little benefit.

For simplicity of discussion, we leave the detailed derivation of the transduction efficiency for Appendix \ref{AppTransEfficiency} (also described in \cite{blesin2021quantum}). Starting from \eqref{transducerhamiltonian}, we first consolidate the equations of motion for the microwave and acoustic fields into one that governs the piezoelectric interaction. Next, we linearize the optomechanical interaction on the assumption that the driving pump is bright enough that it is not noticeably depleted by transduction events. Following this, we transform to the frequency basis to convert the set of linear differential equations into linear algebraic equations, and use the mathematics of signal flow graphs to concisely obtain an amplitude for transduction.

Defining the susceptibilities:
\begin{subequations}
\begin{equation}
\chi_{m}(\omega)=\frac{1}{-i(\omega-\omega_{m}) + \frac{\gamma_{m}}{2}}
\end{equation}
\begin{equation}
\chi_{01}(\omega)=\frac{1}{-i(\omega\!-\!\Delta_{1}) \!+\! \frac{\kappa_{1}}{2}}; \chi_{02}(\omega)=\frac{1}{-i(\omega\!-\!\Delta_{2}) \!+\! \frac{\kappa_{2}}{2}}
\end{equation}
\end{subequations}
we can express the transduction amplitude as:
\begin{equation}\label{RWAtransductionEfficiency1}
\boxed{\frac{\tilde{a}_{\text{out}}}{\tilde{c}_{\text{in}}}\equiv G_{\tilde{a}_{\text{out}},\tilde{c}_{\text{in}}}\!\!\!=\!\!\frac{\sqrt{\kappa_{ex,2}}\sqrt{\gamma_{ex}}\chi_{01}\chi_{02}\chi_{m}\bar{G}\bar{a}_{1}J}{1+\bar{G}^{2}|\bar{a}_{1}|^{2}\chi_{01}\chi_{m}+J^{2}\chi_{01}\chi_{02}}}
\end{equation}
where $\gamma_{m}$ is the overall mechanical linewidth of the bulk acoustic resonator, $\kappa_{1}$ is the resonance linewidth of the first MRR, and $\kappa_{2}$ is the resonance linewidth of the second MRR (which is generally different from $\kappa_{1}$ due to the presence of the coupling to the bus waveguide).

The magnitude square  of $G_{\tilde{a}_{\text{out}},\tilde{c}_{\text{in}}}$ is the microwave-optical transduction efficiency. Here we note that the amplitude for the inverse process of optical-microwave transduction is different due to using $\bar{a}_{1}^{*}$ instead of $\bar{a}_{1}$, but the magnitude square (i.e., the efficiency) is identical in both directions. While this may seem unusual, it is worth pointing out that the forward and reverse processes (i.e., direct vs converse piezoelectricity and photoelasticity vs electrostriction) are coupled with the same coefficients (See Appendix \ref{ThermoSoundApp} for details). 
\newline
\newline
\noindent \emph{Note:} Upon close observation, one might conclude that if light coursing through either MRR should induce vibrations through the optomechanical interaction (i.e., electrostriction), then it is unrealistic to have an optomechanical interaction in the transducer Hamiltonian \eqref{transducerhamiltonian} acting on only the cladding micro-ring resonator. However, assuming that the acoustic modes associated to each MRR are decoupled from one another (e.g., are spatially separated), we may treat the optomechancial coupling in the external resonator as a loss mechanism (incorporated into $\kappa_{0,2}$) that may be incorporated into the Heisenberg-Langevin equations of motion. Moreover, by choice of design, we can minimize this ancillary optomechanical coupling by shaping the substrate around the second MRR so that it sits within a vibrational node at the appropriate mechanical frequency.

In order to use this formula for the transduction efficiency amplitude, we will also need to know $\bar{a}_{1}(\omega)$, which can be solved as a function of the input pump spectrum $\bar{a}_{\text{in}}(\omega)$ using the equations \eqref{meanfields}:
\begin{equation}\label{togetA1}
\boxed{\tilde{\bar{a}}_{1}=\frac{i J\chi_{01}(\omega)\chi_{02}(\omega)\sqrt{\kappa_{ex,2}}}{1+J^{2}\chi_{01}(\omega)\chi_{02}(\omega)}\tilde{\bar{a}}_{\text{in}}}
\end{equation}
In principle, we could substitute this expression into \eqref{RWAtransductionEfficiency1} to have a single overall equation for transduction efficiency, but we refrain from doing so for simplicity.

Where the ratio $\tilde{\bar{a}}_{1}/\tilde{\bar{a}}_{\text{in}}$ is the amplitude gain of the field in the first MRR as a function of frequency (i.e., the amplitude gain spectrum), we will explore how the inter-MRR coupling $J$ causes a splitting in the amplitude gain spectrum in a later section. On resonance, this cavity enhancement factor simplifies to:
\begin{equation}\label{kex2Resonance}
\frac{|\tilde{\bar{a_{1}}}|^{2}}{|\tilde{\bar{a_{\text{in}}}}|^{2}}\approx \frac{64J^{2}\kappa_{ex,2} }{(\kappa_{1}+\kappa_{2})^{2}|(\kappa_{1}-\kappa_{2})^{2}-16J^{2}|}
\end{equation}
at
\begin{equation}\label{resonancesShift}
\omega=\Delta_{1}\pm J\sqrt{1-\frac{\kappa_{1}^{2}+\kappa_{2}^{2}}{8J^{2}}}
\end{equation}
where these resonances are found where the derivative of the cavity enhancement factor with respect to $\omega$ approaches zero.

For this system, $\kappa_{1}^{2}+\kappa_{2}^{2}\ll 8J^{2}$, and we may approximate these resonances at $\omega=\Delta_{1}\pm J$ with FWHM in angular frequency of approximately $(\kappa_{1}+\kappa_{2})/2$.


In the next section, we will describe what goes into determining the piezoelectric interaction constant $g_{EM}$ (which in turn determines $\gamma_{ex}$), the optomechanical interaction constant $\bar{G}$, the inter-MRR evanescent optical coupling constant $J$, and the evanescent bus waveguide coupling constant $\kappa_{ex,2}$.

\section{Physical foundations of the interaction constants}
\subsection{The Piezoelectric interaction}
\subsubsection{Constitutive equations of piezoelectricity}
The equations governing the piezoelectric interaction \cite{osterberg1937piezodielectric,1998PiezoTheory} are determined from thermodynamics. For a full discussion of their derivation, see Appendix \ref{ThermoSoundApp}.

Using standard definitions for the inverse dielectric permittivity tensor $\eta_{ij}$, the elasticity or stiffness tensor $c_{ijk\ell}$, and the stress-voltage form of the piezoelectric tensor $h_{ijk}$ \footnote{Note that the more common stress-charge form of the piezoelectric tensor is $e_{ijk}$ such that $h_{ijk}=\eta_{im}e_{mjk}$.}, and having assumed a negligible difference between the isothermal and isentropic forms of these constants \cite{IEEEPiezo1949}, we obtain the constitutive equations for the piezoelectric interaction:
\begin{subequations}
\begin{align}
X_{ij}&=c_{ijk\ell}x_{k\ell} +h_{ijk}D_{k}\\
E_{i} &= h_{ijk}x_{jk} + \eta_{ij}D_{j}
\end{align}
\end{subequations}
Here we have neglected the effect of temperature changes to the electric field and to the stress, which would be modeled with pyroelectric and thermal expansion tensors (where strain is converted to stress via the elasticity tensor). The elasticity tensor here is defined at a constant electric displacement field. This is known as the piezoelectrically \emph{stiffened} elasticity because the dipole moment built up in response to the applied stress works against it to bring the system back toward equilibrium.

With the constitutive equations of piezoelectricity, and a quantization of linear elasticity (also covered in \cite{blesin2021quantum} and in Appendix \ref{ThermoSoundApp}), one can derive a quantum-mechanical Hamiltonian for the piezoelectric interaction in terms of the quantized electric displacement field $\hat{D}_{i}$ and quantized mechanical displacement operator $\hat{u}_{i}$, which are in terms of the microwave photon and GHz phonon creation/annihilation operators $\hat{c}_{(i,m)}$ and $\hat{b}_{(j,n)}$, respectively:
\begin{align}
\hat{H}_{int}&=-\frac{1}{2}h_{ijk}\int d^{3}r \hat{D}_{i}\frac{\partial}{\partial r_{k}}\hat{u}_{j}\\
&=\sum_{mn}\!\hbar g_{ijk(mn)}\!\!\left(\!\hat{c}_{(\!i,m\!)}(\!t\!)\!+\!\hat{c}_{(\!i,m\!)}^{\dagger}(\!t\!)\!\!\right)\!\!\left(\!\hat{b}_{(\!j,n\!)}(\!t\!)\!+\!\hat{b}_{(\!j,n\!)}^{\dagger}(\!t\!)\!\!\right)
\end{align}
Here, $g_{ijk(mn)}$ is the piezoelectric interaction coupling constant between electromagnetic mode $m$ and acoustic mode $n$:
\begin{equation}
\boxed{g_{ijk(mn)}\!\equiv\! i\frac{\sqrt{\!\frac{\omega_{m}}{\omega_{n}}}}{4V_{\!eff}^{(mn)}}\!\!\sqrt{\!\frac{h_{ijk}^{2}}{\eta_{\text{eff}}^{(m)}\rho_{\text{eff}}}}\!\!\int\!\!\! d^{3}r\mathcal{E}_{(i,m)}\!(\vec{r})\frac{\partial w_{(j,n)}\!(\vec{r})}{\partial r_{k}}}
\end{equation}
Note that this expression subtly differs from that in \cite{blesin2021quantum} due to our quantizing the electromagnetic field in terms of electric displacement field operators, which is needed to make the dynamics consistent with Maxwell's equations \cite{quesada2017you}. In this expression, we have that: $\omega_{m}$ and $\omega_{n}$ are the frequencies of the microwave and mechanical modes of order $m$ and $n$ respectively; $V_{\text{eff}}^{(mn)}$ is the effective electromechanical mode volume for microwave mode $m$ and mechanical mode $n$ (taken to be approximately the volume of the oscillating system); $\eta_{\text{eff}}^{(m)}$ is the effective inverse dielectric permittivity for microwave mode $m$ (approximated as the inverse dielectric permittivity of the material); $h_{ijk}$ is the piezoelectric tensor (in stress-voltage form); $\rho_{\text{eff}}$ is the effective density of the mechanical mode, which we approximate as the density of the mechanical oscillator; $\mathcal{E}_{(i,m)}(\vec{r})$ is the $i^{th}$ component of the microwave spatial mode function of order $m$; and $w_{(j,n)}(\vec{r})$ is the $j^{th}$ component of the mechanical displacement spatial mode function of order $n$, so that its derivative with respect to $r_{k}$ is an expression of the strain in this mechanical mode. That the strain tensor can be described as a single derivative of mechanical displacement works on the assumption that there is no rigid body rotation or torsion within the material (as shown in Appendix \ref{appStraingradient}).

Comparing the piezoelectric coupling tensor $g_{ijk(mn)}$ to our original hamiltonian for the total system and its simplified equations of motion, we have:
\begin{equation}
\sqrt{\gamma_{ex}}\approx -\frac{2 i g_{EM}\sqrt{\Gamma_{ex}}}{\Gamma}
\end{equation}
where $\Gamma$  is the total microwave linewidth of the MRR, equal to the sum of the intrinsic microwave linewidth $\Gamma_{0}$, and the external microwave coupling rate $\Gamma_{ex}$. Where $g_{EM}=i g_{ijk(mn)}$, we obtain a real value for $\sqrt{\gamma_{ex}}$. In general, $\gamma_{ex}$ is a measured quantity rather than derived, but we now see how changing the piezoelectric coupling $g_{EM}$ affects it.

\subsubsection{Analysis of Piezoelectric coupling}
The piezoelectric coupling coefficient $g_{ijk(mn)}$ depends both on bulk material properties of the piezoelectric medium as well as properties that can be varied by design and manufacture.

The bulk properties affecting the piezoelectric coupling include the piezoelectric tensor $h_{ijk}$ (more commonly tabulated as $e_{ijk}=\epsilon_{i\ell}h_{\ell jk}$  so that $h_{ijk}=\eta_{im}e_{mjk}$ \cite{IEEEPiezo1949}) as well as mass density of the material $\rho_{\text{eff}}^{(n)}$ and the effective dielectric constant of the material $\epsilon_{\text{eff}}^{(m)}$ at microwave frequencies.

Apart from having a piezoelectric tensor with large elements, one may increase the strength of the piezoelectric interaction by considering materials of lower density and dielectric constant. However, the molecular origin of the piezoelectric effect is such that these properties are interconnected. For example, a medium of the same atomic polarizeability, but larger number density of atoms will have a larger dielectric constant. The dependence of the dielectric constant on mass density for the same atomic polarizeability will depend on the molar mass of the material.

In general, selecting a bulk material for optimum piezoelectric coupling can be done by considering the ratio $(e_{ijk})^{2}/(\epsilon_{\text{eff}}\rho_{\text{eff}})$ (alternatively, $h_{ijk}^{2}\epsilon_{\text{eff}}/\rho_{\text{eff}}$) as a whole, rather than attempting to control terms individually. See Table \ref{Electromechtable} in Appendix \ref{appa} for comparisons of different materials. Note here, that the dielectric constant is taken at the GHz-scale frequencies, which will generally be substantially larger than what is seen at optical frequencies. For example, Barium Titanate has a tabulated dielectric constant of approximately $2000$, but this does not correspond to a huge refractive index in the telecom band (of order $n=45$). Instead, it is understood that the dielectric constant generally decreases with frequency until leveling off at unity at frequencies so high that the material cannot respond fast enough to the incoming light (i.e., doesn't have enough time to be fully polarized before the direction of the electric field changes).

The adjustable properties of the piezoelectric medium include its overall volume, its particular shape, and the locations of the electrodes all of which may be designed to match mechanical resonances with electromagnetic ones.

The portion of the piezoelectric coupling constant given by these properties is:
\begin{equation}
g_{ijk(mn)}\!\propto\! \frac{h_{ijk}}{\sqrt{V_{\text{eff}}^{EM(m)}V_{\text{eff}}^{mech(n)}}}\!\!\int\!\!\! d^{3}r\mathcal{E}_{(i,m)}\!(\vec{r})\frac{\partial w_{(j,n)}\!(\vec{r})}{\partial r_{k}}
\end{equation}
Scaling up or down the overall volume of the piezoelectric medium should not substantially effect $g_{ijk(mn)}$ because of how the mode volumes are contained in the normalization of $\mathcal{E}_{(i,m)}\!(\vec{r})$ and $w_{(j,n)}\!(\vec{r})$. If we assume that doubling the volume of the medium also doubles the effective mode volumes, then doing so does not change $g_{ijk(mn)}$; the values of $\mathcal{E}_{(i,m)}\!(\vec{r})$ and $w_{(j,n)}\!(\vec{r})$ would each increase by a factor of $\sqrt{2}$, which would cancel out the decrease by the terms outside the integral. However, it is worth noting that one cannot change the dimensions of the medium in isolation. Other parameters, such as the resonant frequencies, would change as well.

While changing the volume (i.e. amount) of the material may not critically improve the piezoelectric coupling, designing the shape of the medium to maximize the integral is an independent concern. In short, one must maximize the overlap of the two functions inside the integral. If the electromagnetic and acoustic modes do not overlap at all, then their product will be zero inside the integral, giving a zero value for the coupling constant. 

One way this zero overlap can occur is when the polarizations of the electric field  $\hat{E}_{i}(\vec{r})$ and the strain-induced electric field $h_{ijk}x_{jk}$ are orthogonal. Whether or not this can be achieved will depend on how the crystal class determines the piezoelectric tensor. Among the tabulated materials in the appendix with a particularly strong piezoelectric effect (i.e.,AlN, $\text{BaTiO}_{\text{3}}$, KTP, LBO, and ZnO), these happen to fall into crystal classes where there are only five distinct nonzero elements to the piezoelectric tensor (though symmetries between elements vary between classes). For these particularly piezoelectric materials, the strain-induced (piezo)electric field would have components $E_{1}=h_{15}x_{13}$; $E_{2}=h_{15}x_{23}$ and $E_{3}=h_{31}x_{11}+h_{32}x_{22} + h_{33}x_{33}$. Although somewhat complicated, we can see that normal strains affect only the $z$ component of the generated electric field, and shear strains affect only the transverse components of the electric field. Conversely, only the transverse components of the electric field can generate shear strain, and only the $z$ component of the electric field can generate normal strain. When designing the piezoelectric component of the transducer, and how it will interface with the optomechanical elements, this geometry must be taken into account. 

Independent of polarization, the overlap integral can be interpreted as an inner product between two functions in position space. For more intuition, we can also consider examining this overlap in momentum space. From this we get the intuition that we can maximize the coupling when the wavelengths of the microwave light and the acoustic vibrations are equal (relative to the momentum uncertainty imposed by the dimensions of the medium). That said, the dimensions of the nonlinear medium impose lower limits to the momentum bandwidth of the acoustic waves (via the uncertainty principle), so that a physically smaller medium will accommodate a larger range of wavelengths over which its acoustic oscillation will be resonant with the microwave input.

\subsection{The Optomechanical interaction}
The optomechanical interaction we consider here is the photoelastic interaction in concert with its converse, electrostriction \cite{osterberg1937piezodielectric}. While the photoelastic effect describes how much the linear-optical constants of a material change per unit strain (e.g, stress-induced birefringence), electrostriction describes the stress generated in a mechanical material per unit optical intensity. Here we note that other effects contribute to this optomechanical effect (e.g., a moving-boundaries effect), but that these effects may be incorporated into an overall \emph{effective} photoelasticity in the same way as the boundary walls of an optical waveguide's effect on propagating light can be incorporated into an effective refractive index.

The strength of the photoelastic effect is parameterized by the photoelastic tensor $p_{ijk\ell}$, which relates the strain $x_{k\ell}$ to the change in the \textcolor{black}{(relative)} inverse permittivity $\epsilon_{0}\eta_{ij}$:
 \begin{equation}
 p_{ijk\ell}\equiv\textcolor{black}{\epsilon_{0}}\left(\frac{\partial\eta_{ij}}{\partial x_{k\ell}}\right)_{T,D,\bar{x}}
 \end{equation}
 The subscripts indicate that we are using the form of the photoelastic tensor at constant temperature, displacement field, and for all other components of strain. Using the same kind of thermodynamic relations connecting the forward and converse piezoelectric effects, one can show that the photoelastic tensor is connected to electrostriction:
 \begin{equation}
p_{ijk\ell}=\textcolor{black}{\epsilon_{0}}\left(\frac{\partial^{2} X_{k\ell}}{\partial D_{i}\partial D_{j}}\right)_{x,T,\bar{D}}.
\end{equation}
See Appendix \ref{appc} for details. Since the displacement and electric fields are directly proportional to one another in this regime, the photoelastic tensor is directly related to the amount of applied stress generated as a quadratic function of the electric field (i.e., electrostriction). 


Although the photoelastic interaction is the basis for the scattering of light by sound waves (i.e., Brillouin scattering), it is worth mentioning that light can be converted into phonons by other means. Where \emph{acoustic} phonons are described by quantizing the vibrations of the material that come from fluctuations in the strain $x_{ij}$, there are also \emph{optical} phonons describing vibrational modes in the medium that do not contribute significantly to an overall strain. These optical phonon modes occur between atoms within a crystal unit cell, vibrating against each other such that the center of mass of each unit cell is approximately constant (which implies the spacing between unit cells and therefore the overall strain are approximately unchanging as well) \footnote{See \cite{kittel1986introduction} for basic model describing optical phonons in a 1D atomic lattice with a two-atom unit cell, and see \cite{cline2017variational} for a thorough discussion of coupled oscillators necessary to extend optical phonons to multi-atom multi-dimensional unit cells.} (See Appendix \ref{appb} for a detailed discussion of this point). Scattering of light off of \emph{optical} phonon modes is known as Raman scattering \cite{agrawalnonlinearFiber}, and does not contribute toward optimizing the transduction efficiency other than as a source of noise.

\subsubsection{The Optomechanical Interaction Hamiltonian}

When a transparent medium is put under strain, it can undergo a shift in its optical properties by the photoelastic effect. When an optical cavity does this, its resonant frequencies will shift in response to the changing permittivity. This shift is well-described to first-order by the modified Bethe-Schwinger formula \cite{yang2015simple,johnson2002perturbation}:
\begin{equation}\label{BetheForm}
\Delta\tilde{\omega}\approx -\tilde{\omega}\frac{\int d^{3}r \left(\Delta \mathbf{\epsilon}_{ij}(\mathbf{r},\tilde{\omega})\tilde{\mathbf{E}}_{i}\tilde{\mathbf{E}}_{j}\right)}{\int d^{3}r \left(\frac{\partial (\tilde{\omega}\epsilon_{ij}(\mathbf{r},\tilde{\omega}))}{\partial \tilde{\omega}}\tilde{\mathbf{E}}_{i}\tilde{\mathbf{E}}_{j}-\frac{\partial (\tilde{\omega}\mu_{ij}(\mathbf{r},\tilde{\omega}))}{\partial \tilde{\omega}}\tilde{\mathbf{H}}_{i}\tilde{\mathbf{H}}_{j}\right)}.
\end{equation}
Although other contributions are present, we can envelop these effects into an effective photoelasticity tensor $p_{ijk\ell}^{\text{eff}}$, in a similar way as boundary effects may result in an effective refractive index in an optical waveguide.

To begin our simplified treatment, the electromagnetic Hamiltonian is given by:
\begin{equation}\label{Ham}
\hat{H}=\frac{1}{2}\int d^{3}r \left(\eta_{ij}\hat{D}_{i}\hat{D}_{j} + \frac{1}{\mu_{ij}}\hat{B}_{i}\hat{B}_{j}\right)
\end{equation}
where the magnetic field modes $\hat{B}_{i}$ are defined relative to the electric displacement field modes $\hat{D}_{i}$ in a manner consistent with Maxwell's equations \cite{quesada2017you}.

Consider the first-order approximation of the effect of strain on the inverse permittivity:
\begin{equation}
\eta_{ij}^{eff}(x_{k\ell})\approx \eta_{ij}(0) + \frac{ p_{ijk\ell}^{(eff)}}{\epsilon_{0}}x_{k\ell}
\end{equation}
If we substitute this expression into the electromagnetic Hamiltonian, we find an interaction term of the form:
\begin{equation}\label{optoMechForm}
\hat{H}_{int}=\frac{1}{2\epsilon_{0}}\int d^{3}r\left( p_{ijk\ell}^{(eff)}\hat{D}_{i}\hat{D}_{j}\hat{x}_{k\ell}\right)
\end{equation}
so that the total hamiltonian when simplified is expressed as:
\begin{align}
\hat{H}&=\sum_{im}\hbar\omega_{m}\hat{a}_{m}^{\dagger}(t)\hat{a}_{m}(t)\\
&\qquad - \sum_{imn}\hbar\bar{G}_{mn}\; \hat{a}_{(i,m)}^{\dagger}(t)\hat{a}_{(i,m)}(t)\left(\hat{b}_{n}(t)+\hat{b}_{n}^{\dagger}(t)\right)\nn
\end{align}
where
\begin{align}\label{GmnOptoMech}
\bar{G}_{mn}\!\!&=\!\!\sqrt{\frac{\hbar}{32\rho_{\text{eff}}^{(n)}V_{\text{eff}}^{mech(n)}\!\epsilon_{0}^{2}(\eta_{\text{eff}}^{(m)})^{2}(\!V_{\text{eff}}^{(m)EM}\!)^{2}\omega_{n}}}  p_{ijk\ell}^{(mn),\text{eff}}\nn\\
&\times\int\!\! d^{3}r \,\mathcal{E}_{(i,m)}(\vec{r})\mathcal{E}^{*}_{(j,m)}(\vec{r})\frac{\partial w_{(k,n)}(\vec{r})}{\partial r_{\ell}}
\end{align}
is the single-photon optomechanical coupling constant. To our knowledge, this is the first expression for the optomechanical coupling in transduction  fully worked out in the literature. Enfolded in the optomechanical coupling constant is the supplementary concern of acoustic impedance matching between the micro-ring resonator, and the bulk medium in which it is embedded. By quantizing the mechanical oscillations of the bulk and (first) MRR together as a single object, the corresponding normal modes automatically include any boundary-related effects between acoustic waves propagating into/out of the MRR.

This optomechanical interaction hamiltonian facilitates frequency shifts related to the value of $\bar{G}_{mn}(\hat{b}_{n}+\hat{b}_{n}^{\dagger})$ between the frequency modes in question. Note here, that because we are considering a single electromagnetic field mode being coupled to the acoustic mode, we do not include the summations over both multiple electromagnetic field modes.

In the transduction scheme considered here, there is a classically bright coherent state pump field in the telecom band driving conversion of GHz-scale phonons into telecom-scale photons. Since the MRRs are being pumped in part by a classically bright driving laser, we may describe the optical field modes as a sum of a classical amplitude and a quantum fluctuation (e.g.,$\hat{a}\rightarrow \bar{a}+\delta\hat{a}$) and linearize the interaction hamiltonian, using the linearization approximation:
\begin{equation}
\hat{a}^{\dagger}\hat{a}\rightarrow (\bar{a}^{*}+\delta\hat{a}^{\dagger})(\bar{a}+\delta\hat{a})\approx |\bar{a}|^{2}+\left(\bar{a}\delta\hat{a}^{\dagger}+\bar{a}^{*}\delta\hat{a}\right)
\end{equation}
(discussed in Appendix \ref{linearizingApp}), which yields the linearized optomechanical interaction hamiltonian:
\begin{equation}\label{OptoMechInter}
\hat{H}_{int}=- \hbar\bar{G}\left(\bar{a}\delta\hat{a}^{\dagger}+\bar{a}^{*}\delta\hat{a}\right)\left(\hat{b} +\hat{b}^{\dagger} \right)
\end{equation}
Here, we have dropped the summation over multiple mechanical and frequency modes. Given a sufficiently narrowband driving pump (relative to the mode spacing), this single-mode approximation is sufficient for studying the transduction efficiency. Where $\bar{G}$ is the single-photon optomechanical coupling constant, $\bar{G}\bar{a}$ is often grouped together as an overall optomechanical coupling constant.

\subsubsection{Analysis of optomechanical coupling}
The most straightforward way to tune the effective optomechanical coupling $\bar{G}\bar{a}$ is to change the power of the pump laser, which changes $\bar{a}$. When this is not reasonable (e.g., it approaches optical damage thresholds), we can see how to change $\bar{G}$ itself both through bulk material properties and design.

Regarding bulk material properties, we see from \eqref{GmnOptoMech} that $\bar{G}_{mn}$ scales linearly with the dielectric constant $\epsilon_{ij}$ (owing to its inverse scaling with $\eta_{ij}$), multiplied by the photoelasticity tensor $p_{ijk\ell}$, and divided by the square root of the density of the material. \textcolor{black}{Unlike the piezoelectric interaction, the dielectric constant here is taken at optical frequencies, and is approximately equal to the square of the refractive index}. In Appendix \ref{appa},  we have tabulated the properties of common optical materials, and one can see for example that barium titanate is a promising candidate for a stronger optomechanical coupling than the silicon nitride used in the MRRs in \cite{blesin2021quantum}. \textcolor{black}{Crystalline silicon also has a higher optomechanical figure of merit than silicon nitride, but its overall optomechanical coupling is limited due to the two-photon absorption that occurs in silicon at higher powers due to its} \textcolor{black}{narrow bandgap.}

For aspects of the optomechanical coupling that can be affected by design, we must look to optimize the overlap integral in $\bar{G}_{mn}$ \eqref{GmnOptoMech}. Starting with polarization, we can focus on how the crystal symmetries affect the photoelasticity tensor, and how that may ultimately affect the optomechanical coupling.
\begin{equation}
\bar{G}_{mn}\propto  \int\!\! d^{3}r \,p_{ijk\ell}^{(mn),eff}\mathcal{E}_{(i,m)}(\vec{r})\mathcal{E}^{*}_{(j,m)}(\vec{r})\frac{\partial w_{(k,n)}(\vec{r})}{\partial r_{\ell}}
\end{equation}

For each crystal class, the symmetries of the photoelastic tensor $p_{ijk\ell}$ are nearly identical to the corresponding symmetries of the (mechanical) elastic tensor $c_{ijk\ell}$ with the exception that $p_{ijk\ell}\neq p_{k\ell ij}$. This means that while the photoelastic tensor can be written as a $6\times 6$ matrix using Voigt notation, it is not guaranteed to also be a symmetric matrix. Knowing this, the symmetries in the respective upper and lower triangular parts of the photoelastic tensor are inherited individually from the corresponding symmetries in the mechanical elastic tensor. We point this out due to the relative abundance of information on the symmetries of the mechanical elastic tensor.

With the symmetries of the photoelastic tensor in mind, and because we have only one optical pump field, we need only consider components in the form of $p_{ijk\ell}$ where $i=j$ when trying to optimize the optomechanical coupling with regards to polarization. For most of the crystal classes of tabulated materials with significant photoelasticity, we find that the significant elements of the photoelasticity tensor are the ones that connect components of normal strain to components of the electric field. In other words, we generally need to consider only the underlying $3\times 3$ matrix $p_{iijj}$. 

Where the piezoelectric and optomechanical media are mechanically bound to one another, the sound waves generated by the piezoelectric medium will dictate what strains need be considered in the optomechanical coupling component. For example, where the electrodes in the transducer are placed to generate an oscillating electric field in the vertical ($z$) direction through the piezoelectric medium, we know that this predominantly produces normal strains $x_{zz}$ also propagating in the $z$ direction. Thus, the elements to the photoelasticity tensor significant to this transducer (in Voigt notation) will be $p_{13}$, $p_{23}$ and $p_{33}$. In Table~\ref{Electromechtable}, we have tabulated values of $p_{33}$, which gauge the extent to which this sound can generate light polarized in the same direction (i.e., TM light polarized out of plane), propagating through the resonator. Where it is more common to fabricate resonators that support single TE (polarized in-plane) optical modes, the important elements of the photoelasticity tensor will be $p_{13}$ and $p_{23}$. Note that because the optical field appears twice in the overlap integral, there is no overall cancellation when integrating over the dimensions of the MRR.

Where conventional integrated photonics employ single-mode waveguides for TE light (polarized in the plane of the medium), MRRs made in this design will have to have maximal $p_{13}$ and $p_{23}$ for maximum optomechanical coupling, though they need not both be large for the coupling to be significant.

To maximize the spatial component of the overlap integral, we use arguments similar to those of phase-matching in nonlinear optics. The MRR is very thin in the $z$-direction, even when compared to the wavelength of the GHz phonons propagating through it. Because of this, we need only ensure that the MRR is located within a vibrational antinode to maximize the strain, as well as its overlap with the optical field.

\subsection{The optical (evanescent) coupling interaction}
One special aspect of the design of this transduction platform is the use of a pair of coupled ring resonators, and using their symmetric and antisymmetric supermodes as the principal optical resonances. A simpler design would be to use just one micro-ring resonator, optomechanically coupled to the piezoelectric medium, and evanescently coupled to the optical bus waveguide. Such a simplified system can in principal have transduction efficiencies comparable to what is seen in the more sophisticated double-ring platform studied here, but at a great cost that compromises its utility. 

The resonances of a single MRR are (nominally) equally spaced in frequency, relative to the integer number of wavelengths occurring in a round trip, and the dispersion in the MRR. When a single resonance is excited with a classically bright pump laser to facilitate transduction between adjacent resonances, one also generates light at the transduced frequency due to (nondegenerate) spontaneous four wave mixing (SFWM), where pairs of generated photons at opposite adjacent resonances are created by annihilating pairs of photons at the pump frequency resonance. This can occur in these systems because the sum of the momenta of a photon one free spectral range (FSR, i.e., the spacing between resonances) below the pump frequency and a photon one FSR above the pump frequency is approximately equal to the sum of the momenta of a pair of photons at the pump frequency, satisfying both energy and momentum conservation (which is commonly called being phase-matched). The relatively large number of photon pairs generated by SFWM, compared to the single photons we seek to obtain as a result of transduction would overwhelm any transduction signal with noise. Moreover, with all resonances more or less equally spaced, and the spacing needing to be only a few GHz apart for microwave-optical transduction, there would be more light generated at many other pairs of resonances due to their being phase-matched, which at high intensity can also feed back into the transduced mode, contributing still more noise photons at the transduction frequency. On top of this, having an MRR large enough to have an FSR in the GHz scale would have a radius of the order of centimeters, making integration challenging at a large scale. 

\begin{figure}[t]
                \centerline{\includegraphics[width=0.95\columnwidth]{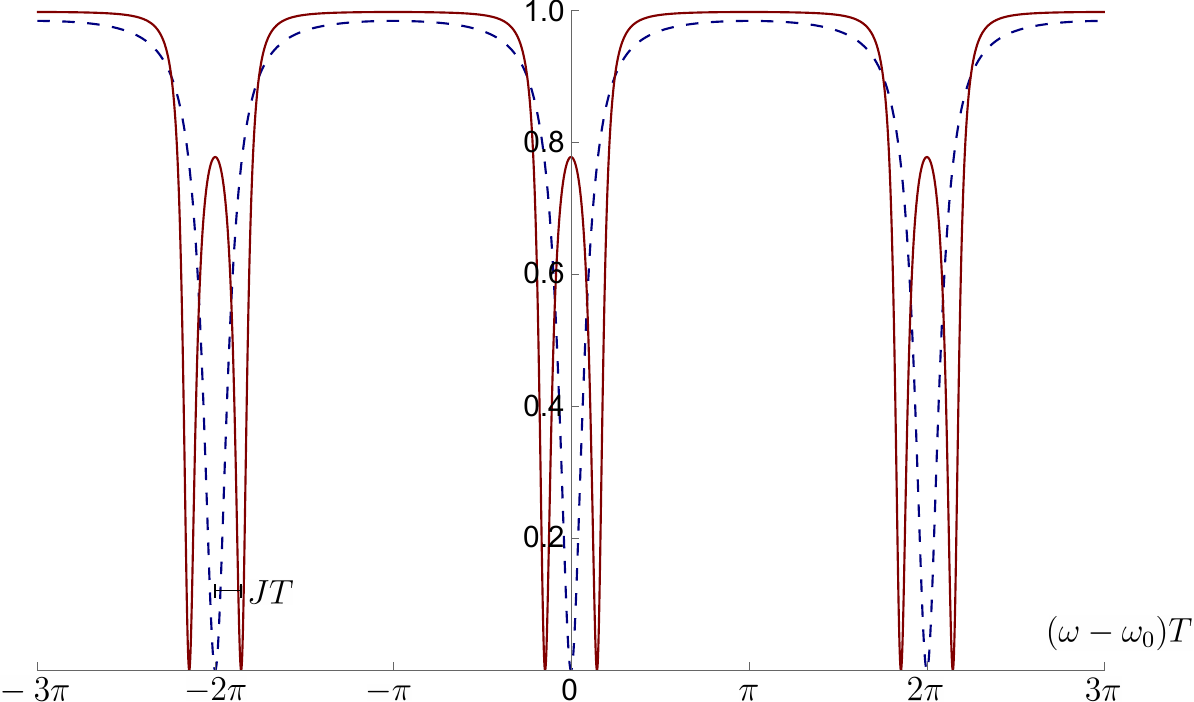}}
\caption{Comparison of transmission spectra through the bus waveguide coupled to either a single ring/double-bus (dashed curve) or a double-ring/double bus (solid curve)  micro ring resonator system. Here, we see that for two identical rings of round-trip time $T$, that the inter-MRR coupling introduces a splitting of the resonances independent of the free-spectral range of the rings. For this plot, we used critical coupling where the round-trip loss in one MRR is equal to the ``reflection" coefficient off of the bus waveguide back into the MRR. Losses and coupling parameters for this plot were chosen to illustrate the structure of the transmission spectrum. By utilizing this splitting, we can use small rings with a large FSR (and therefore large finesse and cavity enhancement factor) whilst keeping the spacing between resonant frequencies of interest in the microwave band to facilitate transduction. Note: if we were to include optical backscattering, this would amount to including a second splitting on top of the already once split peaks (resulting in quadruplets). In assuming the optical backscatter is negligible, we assume this second splitting is too small to be resolved, relative to the linewidth.}\label{RingTransmPlot}
\end{figure}

By utilizing a pair of MRRs for the optical component of the transducer instead of just one, the frequency resonances are similar to the single-MRR case, except that each single peak is split into two adjacent peaks, whose spacing is independent of the round-trip time in the MRRs (See Fig.~\ref{RingTransmPlot} for diagrammatic plot). By making the spacing between the split peaks equal to the microwave frequency, one is free to have a much wider FSR, employing much smaller MRRs (and thus a much smaller footprint) which will consequently have much smaller round trip times, higher finesses, and a larger cavity enhancement factor so that smaller pump powers may be used to drive the system at the same intensity as would be needed in the single-ring system. Because we can choose the FSR to be much larger than the frequency splitting, any light generated by SFWM in peaks equally spaced from the pump resonance can be made so far apart that they are either not phase matched due to dispersion (inhibiting the generation of SFWM light) or else the SFWM light that is generated is still readily filtered out from the transduced light due to the large frequency separation. If we pump one of these resonant peaks with classically bright light, its adjacent peak (from the frequency splitting) will not be well-phase-matched for SFWM since there is no corresponding resonance on the opposite side of the pump resonance with the same frequency separation. Thus, with the double MRR design, one nearly eliminates SFWM as a confounding source of noise at the transduction frequency without compromising transduction efficiency. SFWM is not entirely eliminated, but is greatly suppressed relative to the one-ring case where all wavelengths are resonant. This does not completely eliminate other confounding third-order nonlinear-optical effects, such as the pump intensity effectively changing the index of refraction of the material (known as self-phase modulation, of which electrostriction contributes a small part).

\subsubsection{Inter-MRR coupling constant}
Consider the system of two coupled MRRs, one of which is coupled also to a bus waveguide. For light traveling between the two MRRs, we can define the amplitude transmission coefficient for the light in one MRR to couple into the other resonator as $\sin(JT)$ without loss of generality for some arbitrary $J$. Here, we define $T$ as the round-trip travel time in each MRR (assumed to be the same for both MRRs).

If one were to examine the transmission spectrum of light  passing through the bus waveguide, in the limit of minuscule coupling between the bus waveguide and the MRRs so that the MRRs are approximately a closed system, then one would find that the peaks and valleys of the transmission spectrum are solutions to the equation:
\begin{equation}
(\cos(JT) + \cos(\omega T))\sin(\omega T)=0
\end{equation}
with values:
\begin{equation}
\omega=\Big\{\frac{n\pi}{T},\frac{\pi + 2\pi n}{T}+J,\frac{\pi + 2\pi n}{T}-J\Big\}:n\in \mathbb{Z}
\end{equation}
where $\omega=n\pi/T$ are the critical values where the spectrum is flat between resonances and between split resonances, and $\omega = (\pi+2\pi n)/T \pm J$ are the locations of the split resonances due to coupling between the resonators (as seen in the solid curve in Fig.~\ref{RingTransmPlot}).

If one imagines these transmission peaks at central frequencies $\omega_{0}\pm J$, one can consider a free-field hamiltonian for the light at each of these two frequencies:
\begin{equation}
\hat{H}_{opt}=\hbar(\omega_{0}-J)\hat{a}_{sym}^{\dagger}\hat{a}_{sym} + \hbar(\omega_{0}+J)\hat{a}_{asym}^{\dagger}\hat{a}_{asym}
\end{equation}
The light associated to these modes can be assumed to be traveling back and forth between the two MRR modes $\hat{a}_{1}$ and $\hat{a}_{2}$. If one takes $\hat{a}_{sym}$ to be a symmetric superposition of $\hat{a}_{1}$ and $\hat{a}_{2}$, and $\hat{a}_{asym}$ to be the antisymmetric superposition of $\hat{a}_{1}$ and $\hat{a}_{2}$, i.e., 
\begin{equation}
\hat{a}_{sym}\equiv\frac{\hat{a}_{1}+\hat{a}_{2}}{\sqrt{2}}\qquad:\qquad \hat{a}_{asym}\equiv\frac{\hat{a}_{1}-\hat{a}_{2}}{\sqrt{2}}, 
\end{equation}
then one will find a total optical hamiltonian of the form:
\begin{align}
\hat{H}_{opt}=\hbar\omega_{0}\Big(\hat{a}_{1}^{\dagger}\hat{a}_{1} + \hat{a}_{2}^{\dagger}\hat{a}_{2}\Big) -\hbar J\Big(\hat{a}_{1}^{\dagger}\hat{a}_{2} + \hat{a}_{2}^{\dagger}\hat{a}_{1}\Big)
\end{align}
which is precisely the form describing the system under study here. From this, we may understand that the evanescent optical coupling $J$ between MRRs determines the frequency splitting between the symmetric and antisymmetric supermodes used in this transduction scheme. Tuning this evanescent coupling amounts to changing the spacing between the MRRs, as described in the next subsection.

\subsubsection{waveguide coupling constant}
The waveguide coupling constant $\kappa_{ex,2}$ describes the fraction of evanescent light surrounding the bus waveguide that couples into the MRR next to it and vise versa. To adjust this coupling, one can either change the distance between the bus waveguide and the MRR, or else change the shape of the MRR to change the length over which this interaction between modes occurs.

In general, the optical waveguide coupling is dependent on the wavelength of the light being coupled. If we consider two parallel waveguides close enough to one another that the evanescent field surrounding one waveguide overlaps with the other, we can express the light propagating in the waveguides in a new basis. Instead of separate modes interacting with one another, we can consider superpositions of modes (i.e., supermodes) that are independent of one another. Light initially in one waveguide, is in an even superposition of the symmetric and antisymmetic supermodes, but these modes in general have different mode profiles, and as a result, experience different effective indices of refraction $(n_{\text{eff}}^{(sym)},n_{\text{eff}}^{(asym)})$ throughout the waveguides. Because of this accumulating phase lag between supermodes, the light overall oscillates between waveguides with a characteristic coupling beat length $L_{c}$ \cite{huangCoupledwaveguides1994}:
\begin{equation}
L_{c}\approx\frac{\lambda}{2(n_{\text{eff}}^{(sym)}-n_{\text{eff}}^{(asym)})}
\end{equation}
and the fraction of light in a given waveguide (estimating the magnitude square of the coupling constant) goes as $\cos^{2}(\frac{\pi z}{2 L_{c}})$. 

Note: this means there is also something of a characteristic beat length to the inter-MRR coupling $J$. For maximum bandwidth, we would want the coupling length between the two MRRs to account for no more than one and a half times  this beat length. That said, we must account for the desired value of splitting $J$, which will affect our ability to have minimal wavelength dependence.

In the transducer system considered here, we can understand that if we are considering GHz-scale detunings from a telecom-band wavelength, this is a correspondingly small change in the coupling beat length $L_{c}$, and therefore in the coupling constant. For the small detunings considered in this paper, we may take the waveguide optical coupling to be constant, but still something that may be set at the time of manufacture.

\section{Discussion: Optimizing the transduction efficiency}
In this section, we will describe the basic formula for the total transduction efficiency, and show how it depends in various ways on the different coupling parameters.

Using the equations coupling the different frequency modes in the transducer \eqref{TransducerRWAEquations}, one can derive an overall input-output relation for the transducer, as shown in Section~\ref{MasonGain}:
\begin{subequations}
\begin{align}
\tilde{a}_{\text{out}}(\omega)&=G_{\tilde{a}_{\text{out}},\tilde{c}_{\text{in}}}(\omega)\tilde{c}_{\text{in}}(\omega) + ...\\
G_{\tilde{a}_{\text{out}},\tilde{c}_{\text{in}}}&=\frac{\sqrt{\kappa_{ex,2}}\sqrt{\gamma_{ex}}\chi_{01}\chi_{02}\chi_{m}\bar{G}\bar{a}_{1}J}{1+\bar{G}^{2}|\bar{a}_{1}|^{2}\chi_{01}\chi_{m}+J^{2}\chi_{01}\chi_{02}}
\end{align}
\end{subequations}

 In this section, we will use our results to explore the parameter space of transduction efficiency, and determine how (if possible) one can achieve near unity transduction efficiency. We will use the parameters of the transducer in \cite{blesin2021quantum} as a baseline to see how much systems like these may be optimized.

 \begin{table}[!h]
\begin{center}
\begin{tabular}{|c|c|}
\hline
\multicolumn{2}{|c|}{Nominal Transducer parameters}\\
\hline
\hline
    $\omega_{m}$ & $2\pi\times 3.285$GHz  \\
     \hline
        $\gamma_{0}$ & $2\pi\times 2.6$MHz \\
\hline
$\Gamma_{0}$ & $2\pi\times 500$MHz \\
\hline
$\Gamma$ & $2\pi\times 15$GHz \\
\hline
    $\gamma_{m}$ & $2\pi\times 5.3$MHz \\
    \hline
        $\gamma_{ex}$ & $2\pi\times 2.98$MHz \\
        \hline
        $g_{EM}$ & $2\pi\times 100.6$MHz \\
    \hline
    $J$ & $\pi\times 3.285$GHz \\
\hline
        $\Delta_{1}$ & $2\pi\times 3.285$GHz \\
    \hline
    $\Delta_{2}$ & $2\pi\times 3.285$GHz \\
    \hline
        $\kappa_{1}$ & $2\pi\times 25.$MHz \\
    \hline
        $\kappa_{0,2}$ & $2\pi\times 25.$MHz \\
    \hline
        $\kappa_{ex,2}$ & $2\pi\times 125.$MHz \\
    \hline
        $\bar{G}$ & $2\pi\times 400$Hz \\
    \hline
        $\gamma_{m}$ & $2\pi\times 2.6$MHz \\
    \hline
\end{tabular}
\caption{List of nominal transducer parameters from \cite{blesin2021quantum}.}\label{table2}
\end{center}
\end{table}

\begin{figure}[t]
                \centerline{\includegraphics[width=0.95\columnwidth]{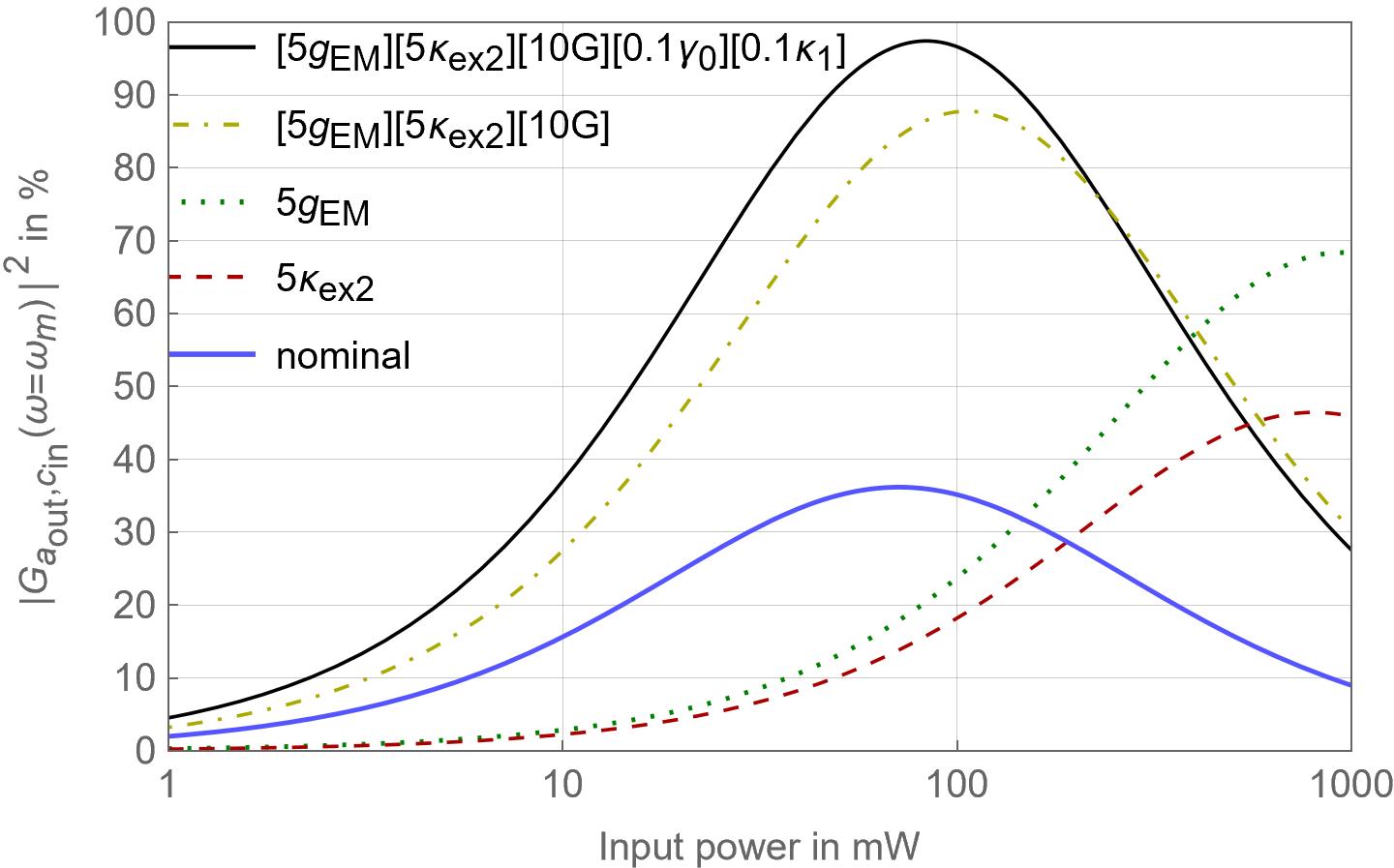}}
\caption{Transduction efficiency \eqref{RWAtransductionEfficiency} as a function of driving laser power on resonance including nominal values from \cite{blesin2021quantum} (solid blue curve); when we increase the external bus waveguide coupling by a factor of $5$ (dashed red curve); when we increase the piezoelecectric coupling by a factor of 5 (dotted green curve); when we increase both these constants by a factor of five, as well as the optomechanical coupling by a factor of ten (dot-dashed yellow curve); and finally, when we increase all of these coupling constants by their aforementioned factors,  as well as lowering the optical and mechanical losses by a factor of ten (solid black curve). Here we see that a maximum transduction efficiency of $97.4 \%$ at $83.6$mW of input power is achievable given suitable material parameters and design.}\label{EffPlotOnResonance}
\end{figure}

Starting with the system parameters in Ref.~\cite{blesin2021quantum}, we can reproduce their efficiency plot as a function of driving laser
power, as shown in Fig.~\ref{EffPlotOnResonance} (shallow solid blue curve).

Where it is not straightforward from inspection to see which values of the system parameters will yield a maximum transduction efficiency \eqref{RWAtransductionEfficiency1}, we can make this analysis much simpler by expressing \eqref{RWAtransductionEfficiency1} in terms of cooperativity parameters and relative coupling efficiency parameters (also called extraction efficiencies in \cite{blesin2021quantum}). Indeed, expressing the transduction efficiency in terms of cooperativities yields simplified transduction efficiency formulas used in the quantum information science community \cite{han2020cavity, jiang2023optically, higginbotham2018harnessing, andrews2014bidirectional, han2021microwave, xie2024scalable, kumar2023quantum}).

The optomechanical cooperativity parameter $\mathcal{C}_{O\!M}$ will be given by the product of two ratios; the ratio of the optomechanical coupling constant $\bar{G}\bar{a}_{1}$ over the optical damping constant $\frac{\kappa_{1}}{2}$, and the ratio of the optomechanical coupling constant $\bar{G}\bar{a}_{1}$ over the mechanical damping constant $\frac{\gamma_{m}}{2}$:
\begin{equation}
\mathcal{C}_{O\!M}\equiv \frac{4\bar{G}^{2}|\bar{a}_{1}|^{2}}{\kappa_{1}\gamma_{m}}
\end{equation}
As a product of ratios of the coupling between two principal modes over their loss/damping constants, cooperativities much greater than unity place us in the regime where the time scale of conversion between the two principal modes is much faster than the time scale of scattering/loss among these modes. Indeed, the square root of cooperativity is the ratio of coupling rate between modes, over the (geometric) mean loss rate per mode.

For simplicity of analysis, we can generalize cooperativity parameters to cooperativity functions (as was done in \cite{blesin2021quantum}) over frequency using the appropriate susceptibilities:
\begin{equation}
\mathcal{C}_{O\!M}(\omega)= \bar{G}^{2}|\bar{a}_{1}|^{2}\chi_{01}(\omega)\chi_{m}(\omega)
\end{equation}
where the original cooperativity parameter is recovered on resonance.

In addition to the optomechanical coupling, we define the inter-ring optical cooperativity $\mathcal{C}_{12}$ as the cooperativity between the respective optical modes in each ring (i.e., between $\hat{a}_{1}$ and $\hat{a}_{2}$) as the corresponding product of ratios of coupling over loss:
\begin{subequations}
    \begin{align}
        \mathcal{C}_{12}&\equiv \frac{4 J^{2}}{\kappa_{1}\kappa_{2}}\\
        \mathcal{C}_{12}(\omega)&= J^{2}\chi_{01}(\omega)\chi_{02}(\omega)
    \end{align}
\end{subequations}
With these cooperativity functions, the expression for the transduction efficiency \eqref{RWAtransductionEfficiency1} greatly simplifies:
\begin{equation}
|\!G_{\tilde{a}_{\text{out}},\tilde{c}_{\text{in}}}\!|^{2}\!\!=\!\frac{\kappa_{ex,2}\!\chi_{02}(\omega)\gamma_{ex}\chi_{m}\!(\omega)}{4} \frac{4\mathcal{C}_{O\!M}(\omega)\mathcal{C}_{12}(\omega)}{(1\!+\!\mathcal{C}_{O\!M}(\omega) \!+\! \mathcal{C}_{12}(\omega))^{2}}
\end{equation}
On resonance, we can simplify the prefactors in the first term of the transduction efficiency as a product of relative coupling efficiencies $\mathscr{F}_{2}$ and $\mathscr{F}_{m}$ (also called extraction efficiencies as they represent the fraction of outcoupling not from intrinsic losses):
\begin{subequations}
\begin{align}
\frac{\kappa_{ex,2}\chi_{02}(\omega)}{2}&\equiv\mathscr{F}_{2}(\omega)\rightarrow \frac{\kappa_{ex,2}}{\kappa_{2}}\\ \frac{\gamma_{ex}\chi_{m}(\omega)}{2}&\equiv\mathscr{F}_{m}(\omega)\rightarrow\frac{\gamma_{ex}}{\gamma_{m}},
\end{align}
\end{subequations}
yielding a transduction efficiency formula on resonance made of independent parameters that is also straightforward to optimize:
\begin{equation}\label{CoopEffic}
|\!G_{\tilde{a}_{\text{out}},\tilde{c}_{\text{in}}}\!|^{2}\!\!\rightarrow\!\mathscr{F}_{2}\mathscr{F}_{m} \frac{4\mathcal{C}_{O\!M}\mathcal{C}_{12}}{(1\!+\!\mathcal{C}_{O\!M} \!+\! \mathcal{C}_{12})^{2}}.
\end{equation}
The portion of the transduction efficiency expression excluding $\mathscr{F}_{2}$ and $\mathscr{F}_{m}$ is known as the internal conversion efficiency of the transducer.

Note: The parameters $(\mathscr{F}_{2},\mathscr{F}_{m},\mathcal{C}_{O\!M},\mathcal{C}_{12})$ are not wholly independent functions, as for example $\mathscr{F}_{m}$ contains some factors shared by $\mathcal{C}_{O\!M}$. However, they are independently tuneable, since there are enough system parameters that one may arbitrarily tune any one of $(\mathscr{F}_{2},\mathscr{F}_{m},\mathcal{C}_{O\!M},\mathcal{C}_{12})$ while holding the other three constant. Although this efficiency formula and its analysis are given in various forms in multiple references \cite{blesin2021quantum, andrews2014bidirectional, higginbotham2018harnessing, jiang2023optically, han2020cavity}, our treatment gives rigorous optimization conditions to obtain a maximal transduction efficiency for intermediate levels of cooperativity.

\subsection{Tuning the couplings}
\subsubsection{Straightforward optimization in terms of cooperativities}
First, we note that the pump power appears only in the optomechanical coupling, and consequently, only in the optomechanical cooperativity $\mathcal{C}_{O\!M}$. This implies that one may reduce the required input pump power for optimum transduction by increasing the single-photon optomechanical coupling $G$. For the transducer in \cite{blesin2021quantum}, $G$ is given in Table \ref{table2} as $2\pi\times 400$ Hz, but much higher (beyond three orders of magnitude) optomechanical couplings are achievable in similar piezo-optomechanical systems as seen in \cite{Vainsencher_2016, Chan_2011, Chan_2012, Balram:14, balram2016coherent, Jiang:19}. 

Since all four-variables in the transduction efficiency expression (equation \eqref{CoopEffic}) can be varied independently, it is straightforward to show \footnote{When all other variables are held constant, the critical value of $\mathcal{C}_{O\!M}$ that maximizes the transduction efficiency may be found by analyzing the zeroes of the derivative of the transduction efficiency with respect to $\mathcal{C}_{O\!M}$ (as is done in the calculus of smooth functions).} that for any fixed set of values $(\mathscr{F}_{2},\mathscr{F}_{m},\mathcal{C}_{12})$, the value of $\mathcal{C}_{O\!M}$ that maximizes the transduction efficiency is where $\mathcal{C}_{O\!M}=\mathcal{C}_{12}+1$. This accounts for the single maximum seen in the transduction efficiency as a function of pump power plotted in Fig.~\ref{EffPlotOnResonance}. For large cooperativities, this also reproduces the impedance-matching rule of thumb of letting $\mathcal{C}_{O\!M}\approx \mathcal{C}_{12}$ to optimize transduction seen in the literature \cite{andrews2014bidirectional, higginbotham2018harnessing, jiang2023optically, han2020cavity}.

Moreover, the fact that the transduction efficiency peaks at a finite power shows that, all other factors constant, optimizing the transduction efficiency is more subtle than simply increasing the coupling coefficients and decreasing the losses. For a full mathematical treatment optimizing the efficiency of a general quantum transducer employing an arbitrary number of intermediate modes, we recommend the reference \cite{WangPRR2022}.

\begin{figure}[t]
\centerline{\includegraphics[width=\columnwidth]{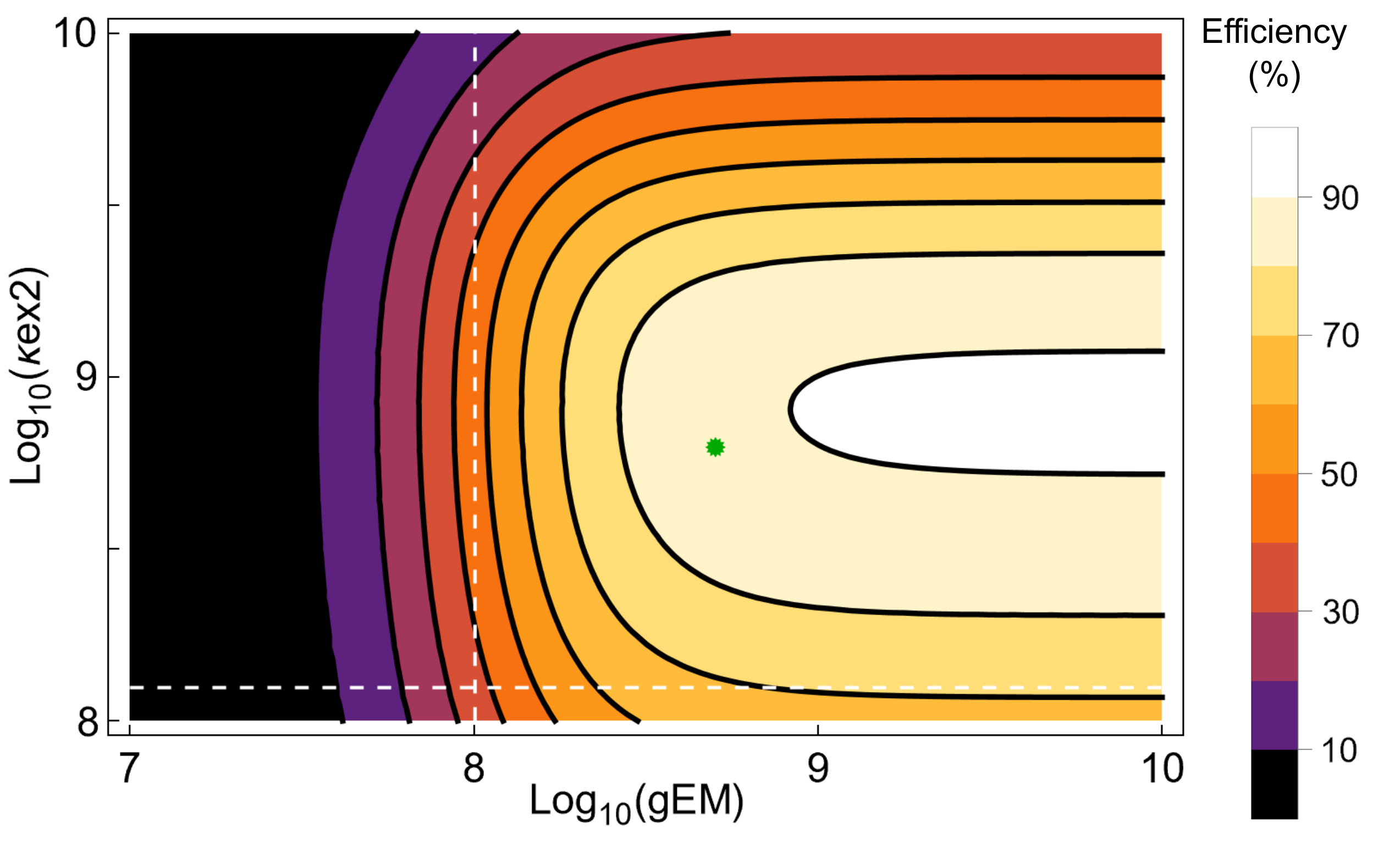}}
\caption{Contour plot of maximum achievable transduction efficiencies (given optimum pump power/optomechanical coupling). The axes are given in the base 10 logarithm of their frequencies (modulo $2\pi$) to smooth out the variation of the efficiency over a large range of parameters. The white dashed crosshairs give the nominal values of $g_{EM}$ and $\kappa_{ex,2}$ for the system studied in this work, corresponding to the peak of the nominal efficiency vs pump power curve in Fig.~\ref{EffPlotOnResonance}. The green pointed dot gives the maximum transduction efficiency when increasing $g_{EM}$ and $\kappa_{ex,2}$ each by a factor of five, as given by the peak of its corresponding curve in Fig.~\ref{EffPlotOnResonance}}\label{MaxxEffPlot}
\end{figure}

In terms of (intra-ring) pump photon number $|\bar{a}_{1}|^{2}$, the critical pump power $|\bar{a}_{1,critical}|^{2}$ to obtain the maximizing value of $\mathcal{C}_{OM}$ is given by the relation:
\begin{equation}
|\bar{a}_{1,critical}|^{2}=\frac{\gamma_{m}}{4\bar{G}^{2}}\left(\frac{4 J^{2}}{\kappa_{2}} +\kappa_{1}\right)
\end{equation}
yielding the maximized efficiency:
\begin{equation}\label{CoopEfficmaxOM}
|\!G_{\tilde{a}_{\text{out}},\tilde{c}_{\text{in}}}\!|^{2}\!\!\rightarrow\!\mathscr{F}_{2}\mathscr{F}_{m} \left(\frac{\mathcal{C}_{12}}{\mathcal{C}_{12}+1}\right).
\end{equation}
With this partially optimized transduction efficiency expressed as a product of independent monotonic functions over the respective domains of independent variables, further maximization is a straightforward selection of whatever system parameters maximize $\mathscr{F}_{2}$, $\mathscr{F}_{m}$, and $\mathcal{C}_{12})$. Moreover, we can directly plot the maximum transduction efficiency as a function of tuning the electromechanical coupling $g_{EM}$, and the waveguide coupling constant $\kappa_{ex,2}$, as shown in Figure \ref{MaxxEffPlot}.

\subsubsection{Optimal system parameters}
Finding maximal values of $\mathscr{F}_{2}$, $\mathscr{F}_{m}$, and $\mathcal{C}_{12}$ can be accomplished a number of ways, such as by minimizing the intrinsic losses within $\kappa_{1}$, $\kappa_{2}$, and $\gamma_{m}$, respectively; or by maximizing the external couplings $\gamma_{ex}$ and $\kappa_{ex,2}$; or some combination of these techniques. 

Where $\gamma_{ex}$ is related to the piezoelectric coupling via the relation:
\begin{align}
\gamma_{ex}&=\left|\frac{2 g_{ijkmn}\sqrt{\Gamma_{ex}}}{\Gamma}\right|^{2},
\end{align}
we understand that selecting materials with higher piezoelectric coefficients (all other factors constant), will increase the maximum transduction efficiency, albeit requiring a higher pump power to do so (see dotted green curve in Fig.~\ref{EffPlotOnResonance}). This is because increasing $\gamma_{ex}$ increases $\mathscr{F}_{m}$, which increases the maximum transduction efficiency. At the same time, increasing $\gamma_{ex}$ decreases $\mathcal{C}_{OM}$, which can be increased back to its optimum value by increasing pump power.

Changing the bus waveguide coupling $\kappa_{ex,2}$ can be accomplished independent of the overall optical coupling of the cladding MRR, $\kappa_{1}$, and the intrinsic optical coupling $\kappa_{0,2}$ (describing optical losses) in the external MRR. Similarly, as shown by the dashed red curve in Fig.~\ref{EffPlotOnResonance}, the maximum transduction efficiency increases with increased waveguide coupling, albeit at a higher power. This can be understood because although increasing $\kappa_{ex,2}$ increases $\mathscr{F}_{2}$, while decreasing $\mathcal{C}_{12}$, the derivative of the transduction efficiency with respect to $\kappa_{ex,2}$ remains positive so long as $\mathscr{F}_{2}$ is below the threshold value $(1+\mathcal{C}_{12})/(2+\mathcal{C}_{12})$, which is above $99.99\%$ for the transducer studied here, making it a straightforward condition to satisfy.

Note: it may seem surprising that one needs higher pump power to recover the optimum transduction efficiency following increasing $\kappa_{ex,2}$ given that a decreasing  $\mathcal{C}_{12}$ implies the optimum value of $\mathcal{C}_{O\!M}=\mathcal{C}_{12}+1$ decreases as well, so that $|\bar{a}_{1}|^{2}$ would actually need to be reduced to attain the maximum efficiency. This all is true, but increasing $\kappa_{ex,2}$ indirectly decreases $\mathcal{C}_{O\!M}$ because it decreases the cavity enhancement factor \eqref{kex2Resonance} between input pump power $|\bar{a}_{in}|^{2}$ and the intra-ring pump power $|\bar{a}_{1}|^{2}$.

By choosing materials with a strong photoelastic response, among other properties, one can also tune the optomechanical coupling rate, though this quantity only enters into the transduction efficiency formula via the driving laser amplitude  \footnote{This is under the assumption that the optomechanical coupling is not a substantial source of loss via electrostriction contributing to $\kappa_{0,2}$.}. Because of this, changing the optomechanical coupling does not change the maximum transduction efficiency, but increasing it does decrease the required laser power needed in order to achieve it.

\textcolor{black}{It is interesting to consider that ultimately, the transduction efficiency asymptotically approaches zero at sufficiently high optical power. One interpretation of this effect would be when the coupling is too large, the transduced light  may be generated either directly via transduction, or as a higher-order effect where part of the transduced light is back-converted and transduced again. These modes of transduced light may interfere destructively with one another in a manner similar to what goes on in optomechanically-induced transparency \cite{Weis2010, Xiong_2018}.}

Taken together, we can see that by simultaneously increasing the optomechanical, electromechanical, and bus waveguide coupling rates (from their nominal values in \cite{blesin2021quantum}), the maximum transduction efficiency (dot-dashed yellow curve in Fig.~\ref{EffPlotOnResonance}) more than doubles its nominal maximum value (solid blue curve in Fig.~\ref{EffPlotOnResonance}) for reasonable experimental parameters. For the system considered here, we point out that there is no one element of the transducer that needs to be made of a material with both a high optomechanical coupling, and a high electromechanical coupling, as these processes occur in different parts of the transducer, and may be optimized independently. In addition, if we decrease the intrinsic losses $\gamma_{0}$, $\kappa_{1}$, and $\kappa_{0,2}$ we directly increase the maximum transduction efficiency by increasing $\mathscr{F}_{m}$, $\mathscr{F}_{2}$, and $\mathcal{C}_{12}$. As an example, the solid black curve in Fig.~\ref{EffPlotOnResonance} shows the maximum transduction efficiency for the system studies here can be increased to above $97\%$ if the intrinsic losses can be reduced by an order of magnitude.

\subsubsection{Mitigation of noise at high efficiency}
At high transduction efficiency, we must answer the question of whether or not the system will be overwhelmed by noise photons in the transduction band washing out the transduced quantum signal. Contributions to the noise arising from sources other than the microwave signal have been discussed comprehensively in \cite{blesin2021quantum}, and will not be reproduced here. However, \textcolor{black}{just as the transduced light generated from the input microwave light is computable as an amplitude, we may compute the amplitudes to produce light at the transduction frequency from every noise source considered in the transducer. To the extent that} the interactions in the transduction process are considered unitary, the sum of the squares of the overall coupling coefficients from each possible input mode \textcolor{black}{(input and noise)} to the output mode $\tilde{a}_{\text{out}}$ must add to unity, so that at higher transduction efficiencies, we see a suppression of the conversion of noise photons into those at the transduced frequency. 

The amount of photons at the transduced frequency due to sources other than the intended input microwave photon will depend on the conditions of the experiments (e.g., thermal noise photons depending on temperature, power instabilities leading to relative intensity noise in the pump, etc). Careful consideration of noise becomes an issue when evaluating the transduction efficiency at the single-photon level, where, say, a cryogenic environment would be required to minimize the number of thermal microwave photons existing in the system that could in principle also be transduced. As current SC qubits operate at cryogenic temperatures, this is not a difficult condition to satisfy in realistic quantum transducers.

\begin{figure}[t]
\centerline{\includegraphics[width=0.95\columnwidth]{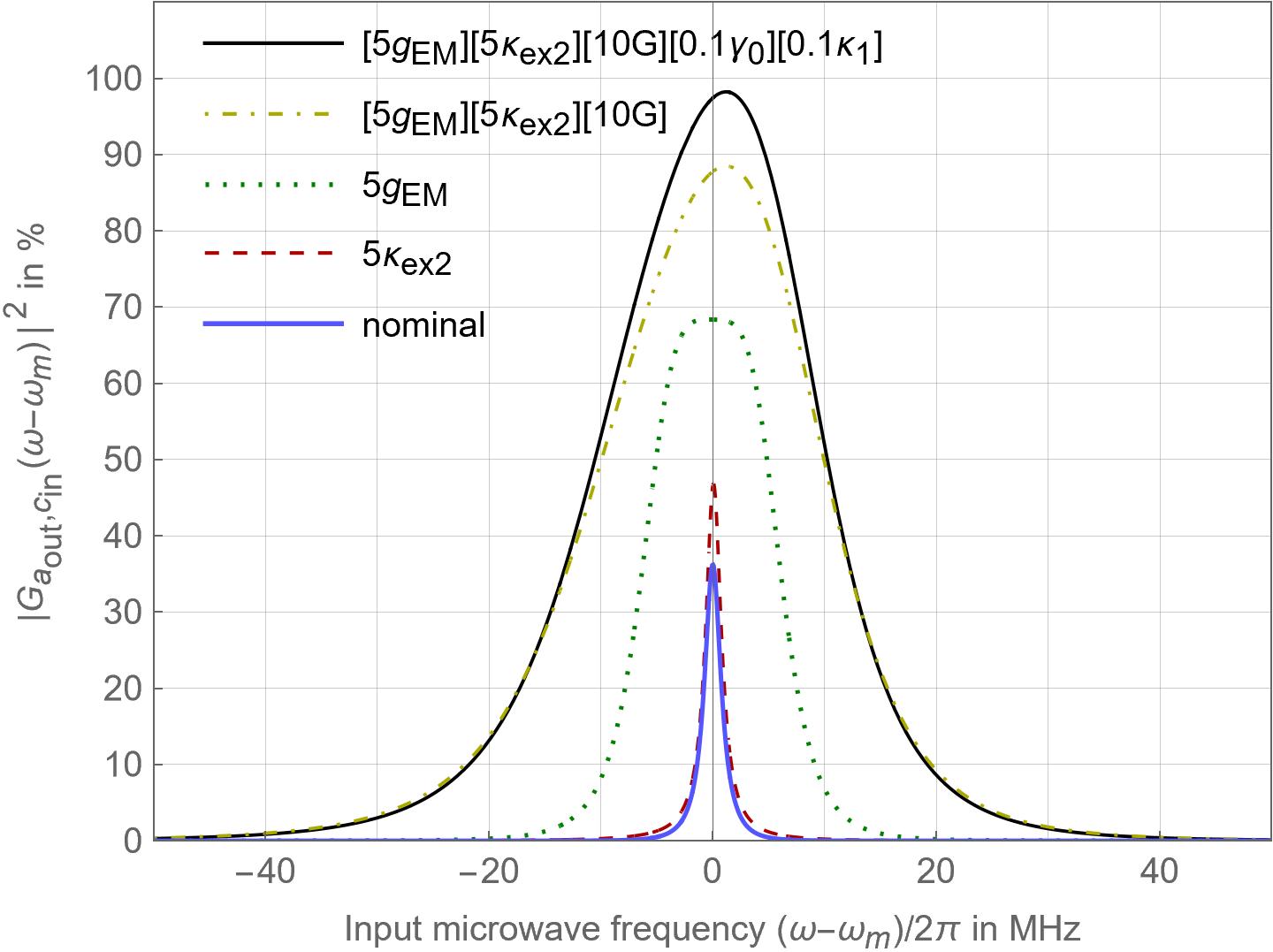}}
\caption{Transduction efficiency \eqref{RWAtransductionEfficiency} as a function of input microwave frequency taken with a driving laser power that maximizes the efficiency for $\omega=\omega_{m}$ for each respective set of parameters, starting with nominal values from \cite{blesin2021quantum}. Here we see that in the nominal case, the bandwidth of near maximal transduction scales is of the order of the intrinsic mechanical linewidth $\gamma_{0}$, but that this bandwidth broadens substantially with increased electromechanical coupling (since $\gamma_{m}$ increases with $g_{EM}$). Moreover, we see a frequency shift appear at large electromechanical and waveguide coupling, which could be due to a combination of the resonance shift of the driving laser, and the frequency shifts arising from linearizing the optomechanical interaction.}\label{EffPlotspectrum}
\end{figure}

\subsection{Transduction efficiency vs bandwidth}
Where our analysis has shown that high transduction efficiencies are achievable at a given microwave frequency, when that frequency is degenerate with the mechanical resonance frequency, it is equally important to consider over what range of frequencies these high efficiencies may be maintained. In Fig.~\ref{EffPlotspectrum}, the transduction efficiency is plotted as a function of the input microwave frequency detuning relative to the mechanical resonance, $\omega-\omega_{m}$. At each frequency, the driving laser power is chosen as to maximize the transduction efficiency on resonance.

With nominal system parameters, the transduction efficiency is nearly uniform over a frequency range comparable to the intrinsic mechanical resonance linewidth. The transduction bandwidth broadens substantially for larger electromechanical coupling, since the total effective mechanical linewidth $\gamma_{m}$ increases with this coupling. In Fig.~\ref{EffPlotspectrum}, we have plotted the transduction efficiency spectra to explore how the transduction bandwidth changes with the coupling parameters.

A surprising feature is that the peaks of these transduction spectra are not all centered on same value, but appear to have a shift that varies with the coupling constants (See Table \ref{ShiftsTable} for shifts in each case). Moreover, the plots in Fig.~\ref{EffPlotspectrum} are taken at the power level that maximizes the peak transduction efficiency on resonance (implicitly setting $\mathcal{C}_{O\!M}(\omega)=\mathcal{C}_{12}(\omega)+1$), which ought to imply that all the transduction spectra have their maximum at the same frequency. However, in our earlier analysis, we had neglected the frequency shift of the mechanical resonance due to the optomechanical interaction when linearizing the optomechanical hamiltonian. In the nominal case, this shift is small compared to the overall transduction bandwidth, but it grows as the coupling rates are increased. A related phenomenon is the resonance frequency shift of the optical cavity, which also increases as the coupling rates are increased, as seen in \eqref{resonancesShift}, but more analysis is needed to fully characterize how the shift varies with the different coupling parameters.

We may compensate for these shifts by changing the frequency of the driving laser, among other tunable parameters. Nevertheless, it is of greater interest that the bandwidth over which the transduction efficiency is high broadens with these increased couplings, so that a broader bandwidth of input states may be transduced with near-unity fidelity as well.

\begin{table}[h!]
\centering
\scriptsize
\begin{tabular}{|c|c|c|}
\hline
case & shift (MHz) & FWHM(MHz)\\
\hline
\hline
nominal & 0.0171 & 1.740\\
\hline
$5\kappa_{ex,2}$ & 0.0799 & 1.546 \\
\hline
$5 g_{EM}$ & (broad peak) & 12.344\\
\hline
$5g_{EM},5\kappa_{ex,2},10G$ & 1.220 & 22.213\\
\hline
$5g_{EM},5\kappa_{ex,2},10G,0.1\gamma_{0},0.1\kappa_{1}$ & 1.207 & 21.366\\
\hline
\end{tabular}
\caption{Frequency shifts and bandwidths using different cases of coupling parameters}\label{ShiftsTable}
\end{table}

\section{Conclusion}
In this article, we have examined the fundamental physical principles behind quantum transduction in a piezo-optomechanical system, and discussed strategies to maximize the transduction efficiency, using the transducer in \cite{blesin2021quantum} as a principal example. To get a better understanding of this concrete system modeled with abstract physics, we focused on the physical foundations of the primary coupling constants in the interaction: the electromechanical coupling constant for the piezoelectric interaction; the optomechanical coupling constant for the optomechanical interaction, and the waveguide coupling constant for the evanescent coupling of light into/out of the transducer. The waveguide coupling constant can be optimized for most materials just by designing the system to have the right spacing and coupling length between the bus waveguide and the second (external) MRR. The electromechanical and optomechanical interaction constants can be maximized both by considering bulk material properties (see Table \ref{Electromechtable} for relevant materials and strengths), and by designing the system to maximize the overlap between the acoustic mode and the electromagnetic mode (or its energy density), respectively.

With these fundamental ideas in mind, we examined how the transduction efficiency changes as a function of these interaction constants using the nominal system parameters in \cite{blesin2021quantum} as a starting point. We found that increasing both the electromechanical coupling and waveguide coupling will increase the maximum transduction efficiency, albeit requiring a substantially higher power for the driving laser. However, by increasing the optomechanical coupling as well, we can decrease the required power of the driving laser to more reasonable values to achieve the maximum efficiency. In achieving these efficiencies, we also examined over what bandwidth we may transduce light at high efficiency, so that we may understand with what fidelity single-photon pulses of finite bandwidth can be transduced. Defining the transduction bandwidth as the range over which the transduction efficiency is greater than $50\%$ of its maximum, we find that this bandwidth can be enhanced substantially by increasing the coupling strength of the electromechanical and waveguide interactions. This comes with a frequency shift that must be accounted for when selecting the correct frequency of the driving laser. Among other things, we also found that it is critical to use a driving laser with linewidth much narrower than the GHz-scale microwave frequencies being transduced in order to avoid two-mode squeezing generating light at the transduction frequency that is not part of the transduction process. With an optimum choice of materials, systems of this type will work well as transducers, provided the light they generate can be well-separated from the driving laser field.

\begin{acknowledgments}
We gratefully acknowledge support from our colleagues at the Air Force Research Laboratory, at Purdue University, and in particular, the insightful comments and discussions with Drs. Michael Senatore and Dylan Heberle, as well as an anonymous referee.

The views expressed are those of the authors and do not reflect the official guidance or position of the United States Government, the Department of Defense or of the United States Air Force. The appearance of external hyperlinks does not constitute endorsement by the United States Department of Defense (DoD) of the linked websites, or of the information, products, or services contained therein. The DoD does not exercise any editorial, security, or other control over the information you may find at these locations.
\end{acknowledgments}

\bibliography{EPRbib16}

\appendix
\begin{widetext}
\newpage

\section{Table of material parameters}\label{appa}
\begin{table}[h!]
\centering
\scriptsize
\begin{tabular}{|c||c|c|c|c|c|c|c|c|}
\hline
 &\multicolumn{5}{|c|}{Material parameters} & \multicolumn{3}{|c|}{Figures of Merit}\\
\hline
\raisebox{-2pt}{Material}                                          & \raisebox{-2pt}{$h_{33}$ ($V/m$)}                                       & \raisebox{-2pt}{$\epsilon_{33}$ ($\text{RF}$)}         & \raisebox{-2pt}{$\epsilon_{33}$ (IR)}     & \raisebox{-2pt}{$\rho$ ($g/cm^{3}$)}     &\raisebox{-2pt}{$p_{3333}\rightarrow p_{33}$}                          &\raisebox{-2pt}{fab.} & \raisebox{-2pt}{[EM]$h_{33}\!\sqrt{\!\frac{\epsilon_{33}(\!\text{RF}\!)}{\rho}}$} & \raisebox{-2pt}{[OM]$\frac{\!\epsilon_{33}\!(\!\text{IR}\!)p_{33}}{\sqrt{\rho}}$}\\[1.5ex]
\hline
$\text{Al}_{2}\text{O}_{3}$ (sapphire)                                     & 0 (c.s.) \cite{GeorgescuSurfacePiezoSapphire2019}                       & 8.6 \cite{wakaki2007physical}                 & 3.05 \cite{querryOptConst}                                      & 3.99 \cite{CRC2015}                      & $-0.2$ \cite{CRC2015}                               & FC                   & 0                                                            & $-3.06\times 10^{-1}$\\
$\text{Al}_{2}\text{O}_{3}$ (alumina)                                      & 0 (c.s.)                                                                & 11.54 \cite{CRC2015}                          & 2.72 \cite{Boidin2016Alumina}                                      & 3.8 \cite{CRC2015}                       & $-0.237$\cite{dragic2013pockels} (clc)              & $\checkmark$         & 0                                                            & $-3.31\times 10^{-1}$\\
\rowcolor{yellowish} $\text{AlN}$                                          & 0.145 \cite{GeorgescuSurfacePiezoSapphire2019}                          & 9.5 \cite{xinjiao1986properties}              & 3.67 \cite{BeliaevAlNIndex}                                      & 3.255 \cite{CRC2015}                     & -0.107 \cite{davydov2002evaluation}                 & $\checkmark$         & $2.47\times 10^{-1}$                                         & $-2.18\times 10^{-1}$\\
$\text{BaB}_{2}\text{O}_{4}$ (BBO)                                         & 0.0224 \cite{guo1989pyroelectric} (clc)                                 & $\approx 7.65$ \cite{trnovcova2007electrical} & 2.37 \cite{Tamosauskas:18}                                      & 3.85 \cite{CRCOpticalMaterialsWeber}     & 0.039 \cite{andrushchak2002photoelastic}            & X                    & $3.16\times 10^{-2}$                                         & $5.39\times 10^{-2}$\\
\rowcolor{aqua} $\text{BaTiO}_{3}$                                         & 0.0017 \cite{GeorgescuSurfacePiezoSapphire2019}                         & 2000 \cite{wakaki2007physical}                & 4.88  \cite{avrahami2003batio}                                    & $\approx 6.00$(m)\!\cite{CRC2015}         & 0.77 \cite{zgonicPRB1994}                           & $\checkmark$         &  $3.15\times 10^{-2}$                                        & $1.53\times 10^{0}$\\ 
$\text{\text{In}}_{2}\text{O}_{3}\!+\!\text{Sn}\text{O}_{2}$(ITO)          & 0 (c.s.) \cite{gonzalezdefect,rodriguez2020study}                       & $\approx 4$ \cite{nagatomo1990electrical}     & $\approx 3.6$ \cite{Moerland:16}                 & 6.8 \cite{ChoiIRODensity2000}            & ?                                                   & $\checkmark$         & 0                                                            & ?\\
$\text{KD}_{2}\text{PO}_{4}$ (DKDP)                                        & 0 (pz) \cite{milek1972potassium,de2015database}                         & 50 \cite{milek1972potassium}                  & (op)\cite{CRCOpticalMaterialsWeber}       & 2.344 \cite{milek1972potassium}          & 0.17 \cite{vantilburg2022piezo}                     & X                    & 0                                                            & (op)\\
$\text{KH}_{2}\text{PO}_{4}$ (KDP)                                         & 0 (pz) \cite{milek1972KDP,de2015database}                               & 44.4 \cite{wakaki2007physical}                & (op)                                      & 2.338 \cite{CRCOpticalMaterialsWeber}    & 0.122 \cite{CRC2015}                                & X                    & 0                                                            & (op)\\
\rowcolor{yellowish} $\text{KTiOPO}_{4}$ (KTP)                             & 0.0935 \cite{ChuKTPPiezoIEEE1992}                                       &  $>17.5$ \cite{Bierlein:89}                   & 3.297 \cite{Kato:02}                      &  3.024 \cite{CRCOpticalMaterialsWeber}   & ?                                                   & FC                   &  $2.25\times 10^{-1}$                                        & ? \\
\rowcolor{yellowish}$\text{KTa}_{1\!-\!x}\text{Nb}_{x}\!\text{O}_{3}$(KTN) & 0.0267 \cite{meng2019elastic}                                           & $\approx 435$ \cite{meng2019elastic}          &$\approx 4.74^{**}$\cite{WANG2023838}                           & 5.759 \cite{meng2019elastic}             & ?                                                   & FC                   & $2.32\times 10^{-1}$                                         & ? \\
$\text{LiB}_{3}\text{O}_{5}$ (LBO)                                         & 0.0971 \cite{wang1995elastic}                                           & 10.71 \cite{kim1997dielectric}                & $\approx 2.5^{***}$\cite{Chen:89}               & 2.474 \cite{CRCOpticalMaterialsWeber}    & 0.3 \cite{zubrinov2004elastic}                      & FC                   & $2.02\times 10^{-1}$                                         & $4.77\times 10^{-1}$\\
$\text{LiNbO}_{3}$                                                         & 0.0433 \cite{GeorgescuSurfacePiezoSapphire2019}                         & $30$ \cite{CRC2015}                           & 4.569 \cite{Zelmon:97}                    & 4.3-4.7 \cite{CRC2015}                   & $0.118$ \cite{mytsyk2021photoelastic}               & FC                   & $1.12\times 10^{-1}$                                         & $2.54\times 10^{-1}$\\
$\text{LiTaO}_{3}$                                                         & 0.0413 \cite{warner1967}                                                & $46$ \cite{CRC2015}                           & 4.506  \cite{BondLiTaO3}                  & 7.45 \cite{CRC2015}                      & -0.044 \cite{CRC2015}                               & FC                   & $1.03\times 10^{-1}$                                         & $-7.26\times 10^{-2}$\\
 $\text{PbZr}_{1-x}\!\text{Ti}_{x}\text{O}_{3}$(PZT)                         & $8.68\!\times\! 10^{-3}\!$(m)\cite{aslam2018analysis}                        & $\approx\! 2200$(m)\cite{Hall1999PZTdielectric}& 5.76 \cite{Ban:21}                                      & 7.5 \cite{sakakibara1994development}     & $?$                                                 & FC                   & $1.49\times 10^{-1}$                                         & $?$\\
$\text{Si}$ (crystal)                                                      & $0$ (c.s.)                                                              & $12.1$* \cite{CRC2015}                        & 12.1 \cite{CRCOpticalMaterialsWeber}                                      & 2.330 \cite{CRC2015}                    & -0.097 (clc) \cite{djemia2013ab}                     & $\checkmark$         & 0                                                            & $-7.69\times 10^{-1}$\\
$\text{Si}$ (amorphous)                                                    & 0                                                                       & $\approx 11.9$ \cite{CRC2015}(clc)            & 12.1 \cite{PhysRevB.5.3017}             & 2.288 \cite{custer1994density}           & ?                                                   & $\checkmark$         & 0                                                            & ? \\
$3C-\text{SiC}$                                                            & 0 (pz) \cite{jain2013commentary}                                        & 9.72 \cite{kimoto2014fundamentals}            & 6.6 \cite{shaffer1971refractive}                                      & 3.21 \cite{kimoto2014fundamentals}       & -0.11 \cite{djemia2013ab}(clc)                      & $\checkmark$         & 0                                                            & $-4.05\times 10^{-1}$\\
$4H-\text{SiC}$                                                            & $-6.04\!\times\! 10^{-3}$ \cite{jain2013commentary}                         & 10.32 \cite{kimoto2014fundamentals}           & 6.81 \cite{wang20134h}                                      & 3.21 \cite{kimoto2014fundamentals}       & ?                                                   & $\checkmark$         & $-1.08\times10^{-2}$                                         & ?\\
$6H-\text{SiC}$                                                            & $-1.16\!\times\! 10^{-2}$ \cite{jain2013commentary}                         & 10.03 \cite{kimoto2014fundamentals}           & 6.57 \cite{wang20134h}                                      & 3.21 \cite{kimoto2014fundamentals}       & ?                                                   & $\checkmark$         & $-2.04\times 10^{-2}$                                        & ?\\
$\text{Si}_{3}\text{N}_{4}$(amorphous)                                     & $0$ (c.s.)                                                              & $4.2$ \cite{CRC2015}                          & 3.99 \cite{luke2015broadband}                                      & $\approx 3.17$ \cite{CRC2015}            & 0.239 \cite{blesin2021quantum}                      & $\checkmark$         & 0                                                            & $5.35\times10^{-1}$\\
$\text{SiO}_{2}$ (amorphous)                                               & 0 (c.s.)                                                                & 3.75\cite{CRC2015}                            & 2.09 \cite{arosa2020refractive}                                      & 2.196 \cite{CRC2015}                     & $0.121$\cite{CRC2015}                               & $\checkmark$         & 0                                                            & $1.45\times 10^{-1}$\\
$\alpha-\text{SiO}_{2}$ (quartz)                                           & 0 \cite{PhysRev.110.1060, grenier2012crystallography, IEEEPiezo1949} (pz) & $4.60$ \cite{CRC2015}                         & 2.36  \cite{ghosh1999dispersion}                                    & 2.648 \cite{CRC2015}                     & 0.1 \cite{CRC2015}                                  & X                    & 0                                                            &  $1.45\times 10^{-1}$\\
$\text{SrTi0}_{3}$ (STO)                                                   &  ?*(pz) \cite{RiMaiSrTiO3TransitionPR1962}                              & $2\times 10^{4}$* \cite{yang2022epitaxial}    & 5.20 \cite{wakaki2007physical}                                      & ?                                        &  ?                                                  & X                    &  ?                                                           &  ?\\
$\beta-\text{Ta}_{2}\text{O}_{5}$                                          & ?                                                                       & 19.5 \cite{nakagawa1989deposition}            & 4.20 \cite{bright2013infrared}                                     & 8.015 \cite{nakagawa1990material}        & $\approx \!-0.05\!$ (est)\cite{nakagawa1993enhancement} & $\checkmark$         & ?                                                            & $-7.41\times 10^{-2}$\\
$\text{ZnO}$                                                               & $0.0906$ \cite{GeorgescuSurfacePiezoSapphire2019}                       & $8.2$ \cite{CRC2015}                          & 3.01 \cite{polyanskiy2024refractiveindex}                                     & 5.6-5.676 \cite{CRC2015}                 & $-0.235$ \cite{CRC2015}                             & X                    & $1.09\times 10^{-1}$                                         & $-2.98\times 10^{-1}$\\
\hline
\end{tabular}
\caption{Electromechanical and photoelastic constants of various materials along with electromechanical [EM] and optomechanical [OM] figures of merit (FOM). Rows tinted in yellow have a large electromechanical FOM, while rows tinted in cyan have a large optomechanical FOM. (c.s.) stands for centrosymmetric. (clc) means the parameter was calculated or estimated from data in the appropriate reference. $(m)$ means the data is taken as the mean value from mutiple differing samples. The symbol $(op)$ means the material is opaque at the particular wavelength considered. The symbol (pz) indicates the material is still piezoelectric, but that the particular coefficient is known to be zero. The symbol (mean) for the piezoelectric coefficient of PZT is because it is the average taken of two different samples labeled PZT in the respective reference. Note: numerous parameters in this table are defined under differing laboratory conditions, but serve to gauge their relative electromechanical coupling. Also, the material PZT has a nonlinear piezoelectric interaction at the high voltages typically used in piezoelectric technologies, but not at the single-photon level considered here. The piezoelectric constants for SiC were estimated using similar polymorphs in \cite{jain2013commentary}. Cells with a question mark were not found in the scientific literature, but are shown to motivate their future determination. (* at cryogenic temperatures) (** averaged over different values of $x$ and taken at 1539nm) (*** extrapolated from Sellmeier equation). Note. for DKDP and KDP, references \cite{milek1972KDP} and \cite{CRCOpticalMaterialsWeber} disagree on the crystal structure (point group), which will affect whether this piezoelectric coefficient $e_{31}$ is necessarily zero. For convenience: $\epsilon_{0}=8.854 187 8128 (13)\times 10^{-12}N/V^{2}$. Note: in \cite{RiMaiSrTiO3TransitionPR1962}, SrTiO$_{3}$ undergoes a phase transition at about 110K, above which, it is centrosymmetric. \textcolor{black}{Note: column labeled Fab. indicates which materials have been successfully used in fabricating integrated photonics (X for no; FC for front-end compatible (can be used in on-chip fabrication only prior to depositing metal electrodes); and $\checkmark$ for yes).}}\label{Electromechtable}
\end{table}

\section{Solving for the transduction efficiency}\label{AppTransEfficiency}
\subsection{Consolidating the Piezoelectric interaction}
The Heisenberg-Langevin equations of motion for the microwave and mechanical fields (not counting optomechanical coupling) are:
\begin{subequations}
\begin{align}
\frac{d\hat{b}}{dt} &= -i\omega_{m}\hat{b}-\frac{\gamma_{0}}{2}\hat{b} - ig_{EM}(\hat{c}+\hat{c}^{\dagger})+\sqrt{\gamma_{0}}\hat{f}_{m}\\
\frac{d\hat{c}}{dt} \!&= \!-i\omega_{MW}\hat{c}\!-\!\frac{\Gamma}{2}\hat{c} \!-\! ig_{EM}(\hat{b}\!+\!\hat{b}^{\dagger})\!+\!\!\sqrt{\Gamma_{0}}\hat{f}_{MW} \!+\!\! \sqrt{\Gamma_{ex}}\hat{c}_{\text{\text{in}}}
\end{align}
\end{subequations}
\textcolor{black}{where $\hat{c}$ is the microwave mode in the cavity (distinct from the transmission line modes $\hat{c}_{\text{\text{in}}}$ and $\hat{c}_{\text{out}}$); where $\gamma_{0}$ is the intrinsic loss rate of the mechanical mode $\hat{b}$, and }where $\hat{f}_{m}$ and $\hat{f}_{MW}$ are (bath) noise-mode annihilation operators for the mechanical and microwave fields, respectively. \textcolor{black}{Note: Because we do not assume any applied \emph{acoustic} input field, we need not define an explicit acoustic input mode $\hat{b}_{\text{\text{in}}}$ as distinct from the mechanical noise mode $\hat{f}_{m}$, when obtaining the Heisenberg-Langevin equation for $\hat{b}$ \cite{gardiner2004quantum, WallsMilburn2008}.}

In frequency space, these Heisenberg-Langevin equations become:
\begin{subequations}
\begin{align}
\Big(-i(\omega & -\omega_{m})+\frac{\gamma_{0}}{2}\Big)\tilde{b} = - ig_{EM}(\tilde{c}+\tilde{c}^{\dagger})+\sqrt{\gamma_{0}}\tilde{f}_{m}\\
\frac{1}{\chi_{MW}}\tilde{c} &= - ig_{EM}(\tilde{b}+\tilde{b}^{\dagger})+\sqrt{\Gamma_{0}}\tilde{f}_{MW} + \sqrt{\Gamma_{ex}}\tilde{c}_{\text{\text{in}}}
\end{align}
\end{subequations}
If we neglect counter-rotating terms, we may substitute the expression for $\tilde{c}$ into the equation of motion for $\tilde{b}$ and (going back to the time domain) obtain the effective equation of motion for $\hat{b}$:
\begin{equation} \label{mechConsPiezoEQM}
\frac{d\hat{b}}{dt}=-i \omega_{m} \hat{b} -\frac{\gamma_{m}}{2}\hat{b} + \sqrt{\gamma_{ex}}\hat{c}_{\text{in}} + \sqrt{\gamma_{noise}} \hat{f}_{piezo}
\end{equation}
where:
\begin{subequations}
\begin{align}
\frac{\gamma_{m}}{2}&=\frac{\gamma_{0}}{2}+g_{EM}^{2}\chi_{MW}\\
\sqrt{\gamma_{noise}}\tilde{f}_{piezo}&= \sqrt{\gamma_{0}}\tilde{f}_{m}-i g_{EM}\chi_{MW}\sqrt{\Gamma_{0}}\tilde{f}_{MW}
\end{align}
\end{subequations}
The fact that we have neglected optomechanical coupling in these equations does not prevent us from making this substitution. The portions of the total system hamiltonian that depend on both $\hat{a}$ and $\hat{b}$ (i.e., the optomechanical interaction) do not depend on $\hat{c}$. Because of this, the equations of motion of the microwave field are the same as an incoming/outgoing free field coupled to a cavity mode $\hat{b}$. Although one could include the optomechanical coupling in the equations of motion of $\hat{b}$, its effect cancels out when deriving the input-output relation for this piezoelectric interaction. In addition, the broad bandwidth implied by the multi-GHz scale of $\Gamma$ in \cite{blesin2021quantum} implies that for the frequencies of order $1$ GHz away from the mechanical resonance, we may still approximate $\chi_{MW}$ as a constant, which makes this overall simplification possible.

\subsection{Piezoelectric and optical input-output relations}
To simplify the equations of motion for the transducer (and from them, obtain the transduction efficiency), we may (as was done in \cite{blesin2021quantum}) interpret the consolidated Heisenberg-Langevin equation of motion for the microwave and mechanical fields \eqref{mechConsPiezoEQM}, as one in which input and output microwave fields ($\hat{c}_{\text{\text{in}}}$, $\hat{c}_{\text{\text{out}}}$) are coupled to the cavity acoustic mode $\hat{b}$ via the input-output relation (in frequency space):
\begin{equation}
\tilde{c}_{\text{\text{in}}}+\tilde{c}_{\text{out}}=\sqrt{\gamma_{ex}}\tilde{b}
\end{equation}
Here, $\gamma_{ex}$ is the external coupling constant between the microwave and mechanical fields. The overall coupling constant $\gamma_{ex}$ is given in terms of the microwave susceptibility $\chi_{MW}(\omega)$, the electromechanical (piezoelectric) coupling rate $g_{EM}$, and the coupling constant $\Gamma_{ex}$ between the microwave input and the microwave mode in the transducer:
\begin{align}
\sqrt{\gamma_{ex}}&=-i g_{EM} \chi_{MW}\sqrt{\Gamma_{ex}}\\
\chi_{MW}&=\frac{1}{-i(\omega\!-\!\omega_{MW}) \!+\! \frac{\Gamma}{2}}\approx \frac{2}{\Gamma}
\end{align}
where the total microwave damping constant $\Gamma$ is the sum of intrinsic microwave loss rate $\Gamma_{0}$ and the loss rate due to (deliberate) external coupling $\Gamma_{ex}$.

In addition to the input-output relation describing the consolidated piezoelectric interaction, we also use an input-output relation to describe the coupling interaction of the pump field in the bus waveguide with the external MRR:
\begin{equation}
\tilde{a}_{\text{in}}+\tilde{a}_{\text{out}}=\sqrt{\kappa_{ex,2}}\hat{a}_{2}
\end{equation}
Here, $\tilde{a}_{\text{in}}$ and $\tilde{a}_{\text{out}}$ are the annihilation operators (in frequency space) for the incoming and outgoing field modes in the bus waveguide, and $\kappa_{ex,2}$ is the optical coupling constant for light moving between the bus waveguide and its adjacent MRR.

To clarify future discussion we assume the pair of MRRs in the transducer are identical (so that $\omega_{1}=\omega_{2}$). Moreover, we assume the pump laser is operating at frequency $\omega_{L}$, and transform the transducer Hamiltonian to a frame of reference rotating with this frequency:
\begin{align}\label{DetunedHamiltonian}
\hat{H} &= \hbar \omega_{m} \hat{b}^{\dagger}\hat{b} +\hbar \Delta_{1} \hat{a}_{1}^{\dagger}\hat{a}_{1} + \hbar \Delta_{2} \hat{a}_{2}^{\dagger}\hat{a}_{2}\nn\\
&\qquad - \hbar J\!\! \left(\!\hat{a}_{1}^{\dagger}\hat{a}_{2} \!+\! \hat{a}_{2}^{\dagger}\hat{a}_{1}\!\right)\nn\\
&\qquad - \hbar \bar{G}\hat{a}_{1}^{\dagger}\hat{a}_{1}\!\! \left(\!\hat{b} \!+\! \hat{b}^{\dagger}\!\right)
\end{align}
where for example, $\Delta_{1}=\omega_{1}-\omega_{L}$. This simplification also makes the equations of motion numerically simpler as all principal rates of oscillation in this frame are of comparable orders of magnitude.

\subsection{Linearizing the optomechanical interaction}\label{linearizingApp}
The second major simplification (also carried out in \cite{blesin2021quantum}) is to linearize the optomechanical interaction by first transforming the hamiltonian with coherent-state displacements of the optical fields, and a corresponding displacement of the mechanical field. For example: $\hat{a}_{1}$ would be transformed to the sum of a mean coherent state amplitude $\bar{a}_{1}$ and the mode annihilation operator $\delta\hat{a}_{1}$ \cite{bowen2015quantum} (so that $\hat{a}_{1}\rightarrow\bar{a}_{1}+\delta\hat{a}_{1}$). After this transformation is carried out, a mechanical displacement is used to simplify the term proportional to $|\bar{a}_{1}|^{2}$ as a negligible zero-point energy term that does not affect system dynamics. Finally, the term proportional to $\delta\hat{a}_{1}^{\dagger}\delta\hat{a}_{1}$ is neglected as insignificant compared to the rest of the system dynamics. Not counting a subtle frequency shift (since we can redefine the resonant frequencies accordingly) and neglecting a separate zero-point energy shift that does not affect the system dynamics, the transducer hamiltonian simplifies to:
\begin{align}\label{transducerhamiltonianv2}
\hat{H} &= \hbar \omega_{m} \hat{b}^{\dagger}\hat{b} +\hbar \Delta_{1} \hat{a}_{1}^{\dagger}\hat{a}_{1} + \hbar \Delta_{2} \hat{a}_{2}^{\dagger}\hat{a}_{2}\nn\\
&\qquad - \hbar J\!\! \left(\!\hat{a}_{1}^{\dagger}\hat{a}_{2} \!+\! \hat{a}_{2}^{\dagger}\hat{a}_{1}\!\right)\nn\\
&\qquad - \hbar \bar{G}\left(\bar{a}_{1}\delta\hat{a}_{1}^{\dagger} +\bar{a}_{1}^{*}\delta\hat{a}_{1}\!\right)\!\! \left(\!\hat{b} \!+\! \hat{b}^{\dagger}\!\right)
\end{align}
For a more comprehensive treatment of this linearization approximation of the optomechanical hamiltonian, we recommend Section 2.7 of Ref.~\cite{bowen2015quantum}.

\subsection{Obtaining the Transducer's equations of motion}
With the hamiltonian linearized, we perform another unitary transformation with respect to the free optical and mechanical hamiltonians so that their quantum operators rotate at their respective frequencies (or detunings). Where the detunings between the central frequencies of $\hat{a}_{1}$ and $\hat{a}_{2}$ and the driving laser $(\Delta_{1},\Delta_{2})$ are of the same order as the GHz-scale vibrational frequencies of the sound being generated by the piezoelectric coupling with microwave photons, we can split the optomechanical interaction neatly into two terms; one driving the transduction, and one driving two-mode squeezing between the acoustic and optical modes:
\begin{align}\label{transducerhamiltonianv3}
\hat{H} &= \hbar \omega_{m} \hat{b}^{\dagger}\hat{b} +\hbar \Delta_{1} \hat{a}_{1}^{\dagger}\hat{a}_{1} + \hbar \Delta_{2} \hat{a}_{2}^{\dagger}\hat{a}_{2}\\
&\qquad - \hbar J\!\! \left(\!\hat{a}_{1}^{\dagger}\hat{a}_{2} \!+\! \hat{a}_{2}^{\dagger}\hat{a}_{1}\!\right)\nn\\
&\qquad - \hbar \bar{G}\left(\bar{a}_{1}\delta\hat{a}_{1}^{\dagger} \hat{b}e^{-i(\Delta_{1}-\omega_{m})t}+\bar{a}_{1}^{*}\delta\hat{a}_{1}\hat{b}^{\dagger}e^{i(\Delta_{1}-\omega_{m})t}\right)\nn\\
&\qquad - \hbar \bar{G} \left(\bar{a}_{1}\delta\hat{a}_{1}^{\dagger}\hat{b}^{\dagger}e^{-i(\omega_{m}+\Delta_{1})t} +\bar{a}_{1}^{*}\delta\hat{a}_{1}\hat{b}e^{i(\omega_{m}+\Delta_{1})t}\right)\nn
\end{align}
Note that these oscillating phases appear because we perform this transformation \emph{after} linearizing the optomechanical interaction.

With the hamiltonian \eqref{transducerhamiltonianv3} partitioned in this way, one can show that if we use a pump laser with frequency $\omega_{L}$ \emph{above} the optical resonance $\omega_{1}$ by amount $\omega_{m}$ (i.e., $\Delta_{1}\approx -\omega_{m}$), we are in the regime of two-mode squeezing since the squeezing portion of the optomechanical interaction oscillates slowly near zero frequency, while the transduction portion oscillates rapidly at frequency $\approx 2\omega_{m}$, and does not contribute significantly to the overall time evolution of the system. Note that this approximation can only be taken if the characteristic interaction time is long enough relative to $1/(2\omega_{m})$, which imposes minimum quality factors that are not very restrictive for the microwave photons, since the full mechanical linewidth is nominally three orders of magnitude smaller than $2\omega_{m}$. However, this does impose a significant restriction to the minimum quality factor of the optical  resonators, with a central frequency multiple orders of magnitude larger than $2\omega_{m}$. It is worth noting that for the nominal transducer parameters in \cite{blesin2021quantum}, that the optical linewidths are still significantly smaller than $2\omega_{m}$, implying a large optical quality factor. In this two-mode squeezing regime, we end up producing entangled phonon-photon pairs of telecom-band photons and GHz-scale phonons instead of converting one into the other. 

Alternatively, if we set the pump laser frequency $\omega_{L}$ to be \emph{below} $\omega_{1}$ by amount $\omega_{m}$ (so that $\Delta_{1}\approx \omega_{m}$), we end up in the transduction regime where the transduction portion of the interaction oscillates slowly, and the squeezing portion oscillates rapidly so that now it is the squeezing that does not contribute significantly to the system dynamics. Considering the system being driven in the two-mode squeezing regime to create entangled microwave-telecom photon pairs has its applications in quantum networking \cite{meesala2023non}, but in this work, we exclusively consider the system in the transduction regime. In this regime, $\omega_{L}\approx \omega_{1}-\omega_{m}$, and we assume the bandwidth of the pump laser is narrow enough that we may neglect any two-mode squeezing contribution to the optomechanical interaction term of the system hamiltonian \eqref{transducerhamiltonianv3}.

If we take this rotating wave approximation, and Fourier-transform to frequency space, then together with the input-output relations, this gives us the following simplified equations of motion for the quantum fields in the transducer:
\begin{subequations}\label{TransducerRWAEquations}
\begin{align}
\frac{1}{\chi_{01}(\omega)}\delta\tilde{a}_{1} &= i J \delta\tilde{a}_{2} + i \bar{G}\bar{a}_{1}(\omega-\Delta_{t})\tilde{b} + \sqrt{\kappa_{0,1}}\tilde{f}_{0,1}\\
\frac{1}{\chi_{02}(\omega)}\delta\tilde{a}_{2}&= i J \delta\tilde{a}_{1} +\sqrt{\kappa_{0,2}}\tilde{f}_{0,2} + \sqrt{\kappa_{ex,2}}\delta\tilde{a}_{\text{in}}\\
\frac{1}{\chi_{m}(\omega)}\tilde{b}&= i \bar{G}\bar{a}_{1}^{*}(\omega+\Delta_{t})\delta\tilde{a}_{1} + \sqrt{\gamma_{0}}\tilde{f}_{m} + \sqrt{\gamma_{ex}}\tilde{c}_{\text{in}}\\
&\tilde{a}_{\text{in}} + \tilde{a}_{\text{out}} = \sqrt{\kappa_{ex,2}}\tilde{a}_{2} \\
&\tilde{c}_{\text{in}}+\tilde{c}_{\text{out}}=\sqrt{\gamma_{ex}}\tilde{b}
\end{align}
\end{subequations}
where $\hat{f}_{0,1}$ and $\hat{f}_{0,2}$ are (intrinsic) noise mode annihilation operators for the optical fields in each resonator; $\kappa_{0,1}$ and $\kappa_{0,2}$ are their corresponding coupling rates; and the transduction detuning $\Delta_{t}=\Delta_{1}-\omega_{m}$ is defined to condense notation. In this way we account for loss due to scattering/absorption as well as due to deliberate out-coupling to the bus waveguide.

Where we have decomposed the optical fields as a sum of a classical mean amplitude and a quantum annihilation operator, and separated out the quantum component to the equations of motion, what remains describes the evolution of the classical mean fields:
\begin{subequations}\label{meanfields}
\begin{equation}
\frac{1}{\chi_{01}(\omega)}\tilde{\bar{a}}_{1} = i J \tilde{\bar{a}}_{2}
\end{equation}
\begin{equation}
\frac{1}{\chi_{02}(\omega)}\tilde{\bar{a}}_{2} = i J \tilde{\bar{a}}_{1} + \sqrt{\kappa_{ex,2}}\tilde{\bar{a}}_{\text{in}}
\end{equation}
\end{subequations}
where we define the susceptibilities:
\begin{subequations}
\begin{equation}
\chi_{m}(\omega)=\frac{1}{-i(\omega-\omega_{m}) + \frac{\gamma_{m}}{2}}
\end{equation}
\begin{equation}
\chi_{01}(\omega)=\frac{1}{-i(\omega\!-\!\Delta_{1}) \!+\! \frac{\kappa_{1}}{2}}; \chi_{02}(\omega)=\frac{1}{-i(\omega\!-\!\Delta_{2}) \!+\! \frac{\kappa_{2}}{2}}
\end{equation}
\end{subequations}
where the shift comes from incorporating the linear phase in the Fourier transform. Note that, from the context of the equations of motion, we have $\kappa_{1}=\kappa_{0,1}$; $\kappa_{2}=\kappa_{0,2}+\kappa_{ex,2}$; and $\gamma_{m}=\gamma_{0} + 4 g_{EM}^{2}/\Gamma$. To simplify things further, we will assume $\Delta_{t}\approx 0$ relative to the laser linewidth. With this, we obtain the simplified equations of motion from which we will obtain the transduction efficiency.

\subsection{Determining the transduction efficiency}\label{MasonGain}
With the equations of motion given as a linear system, we can solve them to obtain $\hat{a}_{\text{out}}$ as a linear function of $\hat{c}_{\text{in}}$, among other variables, and from this, the transduction efficiency.  
\begin{align}
\delta\tilde{a}_{\text{out}} &\approx G_{\tilde{a}_{\text{out}},\tilde{c}_{\text{in}}} \tilde{c}_{\text{in}} + G_{\tilde{a}_{\text{out}},\tilde{a}_{\text{in}}}\delta\tilde{a}_{\text{in}}\nn\\
&+  G_{\tilde{a}_{\text{out}},\tilde{f}_{m}}\tilde{f}_{m}
+G_{\tilde{a}_{\text{out}},\tilde{f}_{0,1}}\tilde{f}_{0,1}\nn\\
&+G_{\tilde{a}_{\text{out}},\tilde{f}_{0,2}}\tilde{f}_{0,2}
\end{align}
Here, the coefficient $G_{\tilde{a}_{\text{out}},\tilde{c}_{\text{in}}}$ is called the overall gain between modes $\tilde{c}_{\text{in}}$ and $\tilde{a}_{\text{out}}$, and is a transfer function in the context of Fourier analysis. As a function of $\omega$, its magnitude square is the transduction efficiency from microwave photons at frequency $\omega$, to optical photons at frequency $\omega + \omega_{L}$.

\begin{figure}[t]
\centerline{\includegraphics[width=0.6\columnwidth]{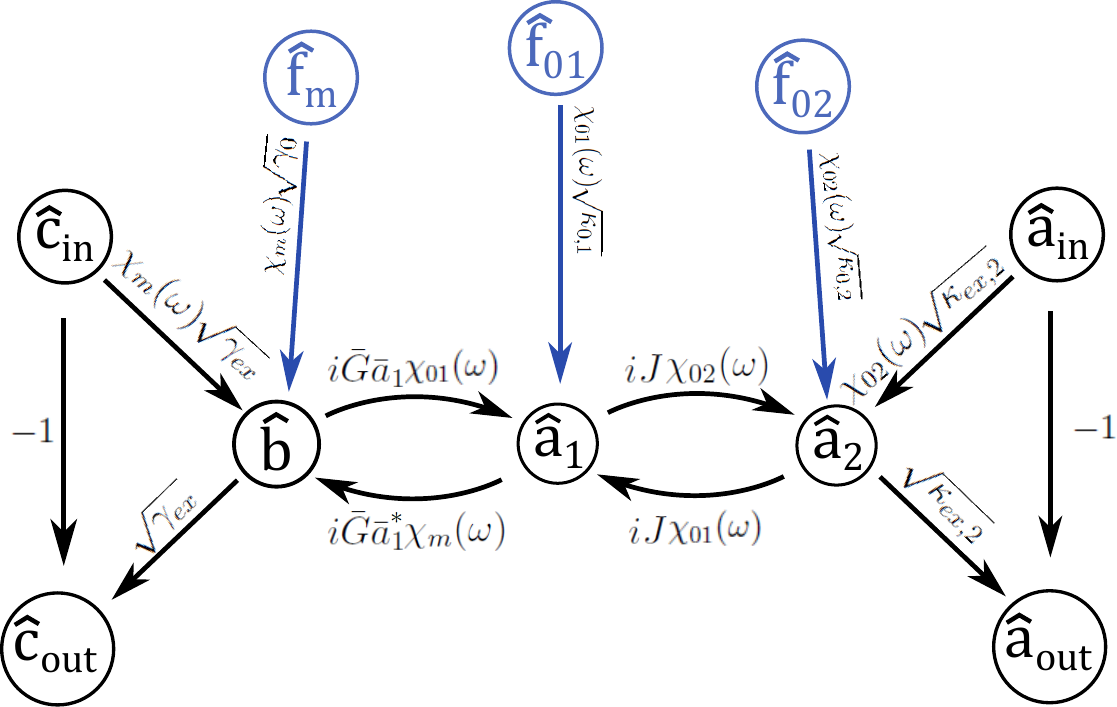}}
\caption{Signal flow graph for the piezo-optomechanical transducer in the rotating wave approximation.}\label{SFGforRWA}
\end{figure}

While one could compute $G_{\tilde{a}_{\text{out}},\tilde{c}_{\text{in}}}$ by brute force from the system of equations \eqref{TransducerRWAEquations}\eqref{meanfields}, reference \cite{blesin2021quantum} introduces representing the system as a signal flow graph and using Mason's Gain rule \cite{mason1956feedback} in order to obtain the transduction efficiency in a simpler fashion that gives intuition on the internal dynamics of the system.

In Fig.~\ref{SFGforRWA}, we show the signal flow graph for our system. With this signal flow graph, we can compute the transduction efficiency as the magnitude square of the overall gain from $\hat{c}_{\text{in}}$ to $\hat{a}_{\text{out}}$.

First, there is only one path from $\hat{c}_{\text{in}}$ to $\hat{a}_{\text{out}}$ with overall gain:
\begin{equation}
G_{1}=\sqrt{\kappa_{ex,2}}\sqrt{\gamma_{ex}}\chi_{01}\chi_{02}\chi_{m}\bar{G}\bar{a}_{1}J
\end{equation}
where the path goes from node $\hat{c}_{\text{in}}$ to $\hat{b}$ to $\hat{a}_{1}$ to $\hat{a}_{2}$ to $\hat{a}_{\text{out}}$ (following along the arrows), and the path gain is the product of the gains between each pair of adjacent nodes in the path.
There are only two loops (which are defined as paths that begin and end on the same node without touching any other node twice), and they have gains:
\begin{align}
L_{1}&=-\bar{G}^{2}|\bar{a}_{1}|^{2}\chi_{01}\chi_{m}\\
L_{2}&=-J^{2}\chi_{01}\chi_{02}
\end{align}
The graph determinant $\Delta$ is defined as $1$, minus the sum of all the single loop gains, plus the sum of all products of pairs of loop gains from non-touching loops (i..e, they share no nodes), minus the sum of all products of triplets of loop gains from non-touching loops, and so on. For a signal flow graph as simple as this transducer, the sum terminates quickly (there are only two loops and they touch each other), and we obtain:
\begin{equation}
\Delta = 1+\bar{G}^{2}|\bar{a}_{1}|^{2}\chi_{01}\chi_{m}+J^{2}\chi_{01}\chi_{02}
\end{equation}
The sub-determinant $\Delta_{1}$ associated to the path with gain $G_{1}$ is defined in the same way as the total graph determinant $\Delta$, except that all loop gains touching this path are set equal to zero. Here, $\Delta_{1}$ is unity because all loops touch this path. 

With this, Mason's Gain formula gives for the overall gain from $\hat{c}_{\text{in}}$ to $\hat{a}_{\text{out}}$:
\begin{equation}
\frac{\tilde{a}_{\text{out}}}{\tilde{c}_{\text{in}}}=\sum_{i}\frac{G_{i}\Delta_{i}}{\Delta}\;\;\rightarrow\;\; \frac{G_{1}\Delta_{1}}{\Delta}
\end{equation}
which in terms of our system variables is given by:
\begin{equation}\label{RWAtransductionEfficiency}
\boxed{\frac{\tilde{a}_{\text{out}}}{\tilde{c}_{\text{in}}}\equiv G_{\tilde{a}_{\text{out}},\tilde{c}_{\text{in}}}\!\!\!=\!\!\frac{\sqrt{\kappa_{ex,2}}\sqrt{\gamma_{ex}}\chi_{01}\chi_{02}\chi_{m}\bar{G}\bar{a}_{1}J}{1+\bar{G}^{2}|\bar{a}_{1}|^{2}\chi_{01}\chi_{m}+J^{2}\chi_{01}\chi_{02}}}
\end{equation}
whose magnitude square is the microwave-optical transduction efficiency. Here we note that the amplitude for the inverse process of optical-microwave transduction is different due to using $\bar{a}_{1}^{*}$ instead of $\bar{a}_{1}$, but the magnitude square (i.e., the efficiency) is identical in both directions. While this may seem unusual, it is worth pointing out that the forward and reverse processes (i.e., direct vs converse piezoelectricity and photoelasticity vs electrostriction) are coupled with the same coefficients. This point will be elaborated on in the next Appendix.
\newline
\newline
\noindent \emph{Note:} Upon close observation, one might conclude that if light coursing through either resonator should induce vibrations through the optomechanical interaction (i.e., electrostriction), then it is unrealistic to have an optomechanical interaction in the transducer Hamiltonian \eqref{transducerhamiltonian} acting on only the cladding micro-ring resonator. However, assuming that the acoustic modes associated to each MRR are decoupled from one another (e.g., are spatially separated), we may treat the optomechancial coupling in the external resonator as a loss mechanism (incorporated into $\kappa_{0,2}$) that may be incorporated into the Heisenberg-Langevin equations of motion. Moreover, by choice of design, we can minimize this ancillary optomechanical coupling by shaping the substrate around the second MRR so that it sits within a vibrational node at the appropriate mechanical frequency.

In order to use this formula for the transduction efficiency amplitude, we will also need to know $\bar{a}_{1}(\omega)$, which can be solved as a function of the input pump spectrum $\bar{a}_{\text{in}}(\omega)$ using the equations \eqref{meanfields}:
\begin{equation}\label{togetA1}
\boxed{\tilde{\bar{a}}_{1}=\frac{i J\chi_{01}(\omega)\chi_{02}(\omega)\sqrt{\kappa_{ex,2}}}{1+J^{2}\chi_{01}(\omega)\chi_{02}(\omega)}\tilde{\bar{a}}_{\text{in}}}
\end{equation}
In principle, we could substitute this expression into \eqref{RWAtransductionEfficiency} to have a single overall equation for transduction efficiency, but we refrain from doing so for simplicity.

\section{Thermodynamic and quantum Foundations of the piezoelectric and optomechanical interactions}\label{ThermoSoundApp}

The differential change in internal energy $dU$ of a dielectric (of which piezoelectric media are a subtype) can be given by the expression:
\begin{equation}\label{thermo1}
dU = TdS +X_{ij}dx_{ij} + E_{i}dD_{i}
\end{equation}
where $TdS$ is the heat flowing in/out of the system, $X_{ij}dx_{ij}$ is the work done on/by the system, and $E_{i}dD_{i}$ is the change of electromagnetic energy in the dielectric. Here, $X_{ij}$ is the mechanical stress tensor describing the various components of forces acting on various planes of the dielectric, while $x_{ij}$ is the mechanical strain tensor, describing the deformation of the medium. To condense notation, we use the Einstein summation convention, where repeated Cartesian indices implies summation over them.

The constitutive equations of the piezoelectric interaction come from the Maxwell relation of either this internal energy, or one of its Legendre transformed free-energy analogues. Using the Helmholtz free energy $F$ such that:
\begin{subequations}
\begin{align}
F&=U-TS\\
dF &= -S dT + X_{ij}dx_{ij} + E_{i}dD_{i}
\end{align}
\end{subequations}
and assuming  that the derivatives in the Maxwell relation are approximately constant for the fields and stresses considered, we obtain:
\begin{subequations}
\begin{equation}\label{maxwell1}
X_{ij} \approx \left(\frac{\partial X_{ij}}{\partial x_{k\ell}}\right)_{T,E,\bar{x}}\!\!x_{k\ell} + \left(\frac{\partial X_{ij}}{\partial D_{k}}\right)_{T,D,\bar{x}}\!\!D_{k}
\end{equation}
\begin{equation}\label{maxwell2}
E_{i} \approx \left(\frac{\partial E_{i}}{\partial x_{jk}}\right)_{T,D,\bar{x}} x_{jk} + \left(\frac{\partial E_{i}}{\partial D_{j}}\right)_{x,T,\bar{D}} D_{j}
\end{equation}
\end{subequations}
This assumption of approximate constancy is particularly well-justified for the single-quantum-level perturbations considered in quantum transduction, but generally applies to macroscopic fields and stresses whenever these interrelations are approximately linear (e.g., we remain in the regime of elastic deformation and linear optics).
Note that the derivative in the second term in \eqref{maxwell1} is equal to the derivative of the first term in \eqref{maxwell2} since these are mixed second-derivatives of the Helmholtz free energy $F$ (i.e., form a Maxwell relation). These constitutive equations would ordinarily each have an additional term containing derivatives of temperature (e.g., pyroelectric/electrocaloric coefficients), but may be neglected assuming either a near-constant, or near-zero temperature. 

Using standard definitions for the inverse dielectric permittivity tensor $\eta_{ij}$, the elasticity or stiffness tensor $c_{ijk\ell}$, and the stress-voltage form of the piezoelectric tensor $h_{ijk}$ (note that the more common stress-charge form is $e_{ijk}$ such that $h_{ijk}=\eta_{im}e_{mjk}$ and we have assumed a negligible difference between isothermal and isentropic forms of these constants \cite{IEEEPiezo1949}) gives us the constitutive equations for the piezoelectric interaction
\begin{subequations}
\begin{align}
X_{ij}&=c_{ijk\ell}x_{k\ell} +h_{ijk}D_{k}\\
E_{i} &= h_{ijk}x_{jk} + \eta_{ij}D_{j}
\end{align}
\end{subequations}
Based on the origin of these constitutive equations, we can be aware that they are isothermal constants defined at some fixed temperature. The elasticity tensor here is defined at a constant electric displacement field. This is known as the piezoelectrically \emph{stiffened} elasticity because the dipole moment built up in response to the applied stress works against it to bring the system back toward equilibrium. This is in contrast to the constant electric field case, where the ends of the piezoelectric material would be in electrical contact so that charge is free to flow to cancel out the generated electric dipole moment. By considering how mechanical energy is converted into electrical energy when a piezoelectric medium under strain is released with and without short-circuiting the system, one can find the electromechanical coupling tensor, but that is beyond the scope of this work.
\newline
\newline
\noindent \emph{Note:} The material-property tensors discussed here with more than two indices are often expressed in the literature with fewer indices. In this case, all distinct components are accounted for, but tabulated using Voigt notation (e.g., $c_{2223}\rightarrow c_{24}$).

\subsection{Quantizing sound for the piezoelectric and optomechanical interactions}\label{AppTheOtherQED}
The framework of quantum optics is based on a canonical quantization of the classical electromagnetic field. Generally speaking, one decomposes the field into a set of orthogonal modes, finds the Lagrangian for the electromagnetic field for each mode, and transforms it to a Hamiltonian, obtaining canonical coordinates and momenta in the process. Where the hamiltonian for these field modes is mathematically equivalent to a set of simple harmonic oscillators, we can replace these conjugate coordinates and momenta with conjugate quantum operators. Where the classical conjugates have a unit Poisson bracket, the conjugate quantum operators  will have a commutator equalling $i\hbar$. In a similar way, we may quantize the acoustic field in an elastic medium.

For an elastic medium, its Lagrangian density is given by the difference between kinetic and potential energy per unit volume:
\begin{equation}\label{lagrangedensity}
\mathcal{L}_{elas}=\frac{1}{2}\rho \dot{u}_{i}\dot{u}_{i} - \frac{1}{2}c_{ijk\ell}x_{ij}x_{k\ell}
\end{equation}
Here, $u_{i}(\vec{r})$ represents the $i^{th}$ component of the displacement of the medium from its original location given by vector $\vec{r}$ and dots over vector components (e.g., $\dot{u}_{i}$) represent time derivatives. In the approximation that we neglect rigid body rotation, we can replace the elements of the strain tensor $x_{ij}$ with corresponding spatial derivatives of displacement $du_{i}/dr_{j}$ \textcolor{black}{(see Appendix \ref{appStraingradient} for details)}\footnote{\textcolor{black}{The assumption of no rigid body rotation at the micro-scale means we will not be considering torsion waves in this system, which couple poorly to both the microwave and optical single modes in this system, but does not preclude them from being useful in other geometries.}}.  This Lagrangian can be transformed to the Hamiltonian:
\begin{equation}
H_{elas}=\frac{1}{2}\int d^{3}r \left[\rho \dot{u}_{i}\dot{u}_{i} + c_{ijk\ell}\frac{\partial u_{i}}{\partial r_{j}}\frac{\partial u_{k}}{\partial r_{\ell}}\right]
\end{equation}
Here, the displacement can be decomposed as a sum over acoustic modes:
\begin{equation}
u_{i}(\vec{r},t)=\sum_{n}\left(u_{n}^{(0)}w_{(i,n)}(\vec{r})e^{-i\omega_{n}t} +u_{n}^{(0)*}w_{(i,n)}^{*}(\vec{r})e^{i\omega_{n}t}\right)
\end{equation}
where $u_{n}^{(0)}$ is an overall amplitude factor scaling between the normalized mode functions $w_{(i,n)}(\vec{r})$ and the total displacement $u_{i}(\vec{r},t)$. \textcolor{black}{Note: we use the comma and parentheses to distinguish the $i^{th}$ component of the $n^{th}$ cavity mode (e.g., $\hat{c}_{(i,n)}$) from the propagating input mode (e.g., $\hat{c}_{\text{in}}$).}

The mode functions themselves $w_{(i,n)}(\vec{r})$ can be found as solutions to the elastic equations of motion:
\begin{equation}
-\omega_{n}^{2}\rho(\vec{r})w_{(i,n)}(\vec{r}) = \sum_{jk\ell}\frac{\partial}{\partial r_{j}}c_{ijk\ell}\frac{\partial w_{(k,n)}(\vec{r})}{\partial r_{\ell}}
\end{equation}
which are also the Euler-Lagrange equations to the lagrangian density \eqref{lagrangedensity}.

The normalization of the acoustic mode functions $w_{(i,n)}(\vec{r})$ can be set by defining an effective mode volume for each mechanical mode:
\begin{equation}\label{veff}
V_{\text{eff}}^{mech(n)}=\Bigg(\int d^{3}r \bigg(\frac{\sum_{i} |w_{(i,n)}(\vec{r})|^{2}}{\int d^{3}r \sum_{i} |w_{(i,n)}(\vec{r})|^{2}}\bigg)^{2}\Bigg)^{-1}.
\end{equation}
In short, we take the fraction of $\sum_{i} |w_{(i,n)}(\vec{r})|^{2}$ divided by its integral over all space as a normalized density (integrating to unity). The average value of this density is equal to the integral of \emph{its square} over all space. The effective mode volume $V_{\text{eff}}^{mech(n)}$ is the volume, which when multiplied by this average density gives unity. With the mechanical mode volume defined, the normalization condition for the acoustic mode functions is given by the convention:
\begin{equation}
\int d^{3}r \sum_{i} |w_{(i,n)}(\vec{r})|^{2}=V_{\text{eff}}^{mech(n)}
\end{equation}
Note that our definition of the mechanical mode volume \eqref{veff} is independent of scale changes to $w_{(i,n)}(\vec{r})$.

Next, one can use the mass density $\rho(\vec{r})$ as a weighting function behind an orthogonality relation for the mechanical mode functions:
\begin{equation}\label{orthofunction}
\int d^{3}r \rho(\vec{r}) \sum_{i} w_{(i,m)}^{*}(\vec{r})w_{(i,n)}(\vec{r})=m_{\text{eff}}^{(n)}\delta_{mn}
\end{equation}
Here, $m_{\text{eff}}^{(n)}$ is the effective mass of mode $n$ of the transducer. We can also use this to define an effective mass density $\rho_{\text{eff}}$ generally equal to the material mass density, such that $\rho_{\text{eff}} V_{\text{eff}}^{(n)}=m_{\text{eff}}^{(n)}$.

Using the orthogonality relation and previous definitions, one can simplify the mechanical hamiltonian greatly:
\begin{align}
H_{elas}&=\sum_{m}m_{\text{eff}}^{(m)}\omega_{m}^{2} (u_{m}^{(0)}u_{m}^{(0)*}+u_{m}^{(0)*}u_{m}^{(0)})
\end{align} 
When applying the quantization procedure, we can use the relations:
\begin{subequations}
\begin{align}
u_{n}^{(0)}e^{-i\omega_{n}t}&\rightarrow \sqrt{\frac{\hbar}{2 m_{\text{eff}}^{(n)}\omega_{n}}}\hat{b}_{n}(t)\\ u_{n}^{(0)*}e^{i\omega_{n}t}&\rightarrow \sqrt{\frac{\hbar}{2  m_{\text{eff}}^{(n)}\omega_{n}}}\hat{b}_{n}^{\dagger}(t)
\end{align}
\end{subequations}
to obtain the quantum elastic Hamiltonian:
\begin{equation}
\hat{H}_{elas}=\sum_{n}\hbar\omega_{n}\left(\hat{b}^{\dagger}_{n}\hat{b}_{n} + \frac{1}{2}\right)
\end{equation}
with displacement operator:
\begin{equation}
\hat{u}_{i}(\vec{r},t) \!=\!\sum_{n}\!\!\sqrt{\frac{\hbar}{2m_{\text{eff}}^{(n)}\omega_{n}}}\left(w_{(i,n)}(\vec{r})\hat{b}_{n}(t) \!+\! w_{(i,n)}^{*}(\vec{r})\hat{b}_{n}^{\dagger}(t)\right)
\end{equation}

The piezoelectric interaction hamiltonian comes from the sum of the quantum elastic hamiltonian and the Electromagnetic  hamiltonian, where the electric field is modified according to the constitutive equation of the piezoelectric interaction:
\begin{equation}
\hat{H}_{EM}=-\int d^{3}r \left(\frac{1}{2}\eta_{ij} \hat{D}_{i}\hat{D}_{j} +\frac{1}{2}h_{ijk}\hat{D}_{i}\hat{x}_{jk}\right).
\end{equation}
where:
\begin{equation}
\hat{x}_{jk}=\frac{\partial \hat{u}_{j}}{\partial r_{k}}
\end{equation}
and as in standard quantum optics in a dielectric, the electric displacement field operator is given by:
\begin{equation}\label{efielddef}
\hat{D}_{i}(\!\vec{r},\!t\!)\!=\! i\!\!\sum_{n}\!\!\sqrt{\!\frac{\hbar\omega_{n}}{2\eta_{eff\!,\!(\!n\!)}\!V_{\text{eff}}^{EM\!(\!n\!)}}}\!\left(\!\!\mathcal{E}_{(\!i,n\!)}\!(\!\vec{r})\hat{c}_{(\!i,n\!)}\!(\!t\!) \!-\! \mathcal{E}_{(\!i,n\!)}^{*}\!(\!\vec{r})\hat{c}_{(\!i,n\!)}^{\dagger}\!(\!t\!)\!\!\right)
\end{equation}
with mode functions $\mathcal{E}_{(i,n)}\!(\!\vec{r})$ normalized by 
\begin{equation}
\int d^{3}r \sum_{ij}\eta_{ij}(\vec{r})\mathcal{E}_{(j,m)}(\vec{r})\mathcal{E}_{(i,n)}^{*}(\vec{r})=\delta_{mn}\eta_{eff,(n)}V_{\text{eff}}^{EM(n)}
\end{equation}
and the electromagnetic mode volume is defined similarly to the acoustic mode volume:
\begin{equation}
V_{\text{eff}}^{EM(n)}=\frac{\left(\int d^{3}r \sum_{ij}\eta_{ij}(\vec{r})\mathcal{E}_{(j,n)}(\vec{r})\mathcal{E}_{(i,n)}^{*}(\vec{r})\right)^{2}}{\int d^{3}r \left(\sum_{ij}\eta_{ij}(\vec{r})\mathcal{E}_{(j,n)}(\vec{r})\mathcal{E}_{(i,n)}^{*}(\vec{r})\right)^{2}}
\end{equation}
Note here, that we separate out the factor of $i$ in the definition of the field modes $\mathcal{E}_{(i,n)}(\vec{r})$. As an example of the properties of the spatial mode function, we may use the quantization procedure in \cite{schneeloch2019introduction} to examine $\mathcal{E}_{(i,n)}(\vec{r})$ in the case of Hermite-Gaussian beam propagation to be:
\begin{equation}
\mathcal{E}_{(i,n)}(\vec{r})\propto \frac{g_{\mu}(x,y)}{\sqrt{\mathscr{L}_{z}}}e^{i k(n) z}(\vec{\epsilon}_{kn})_{i}
\end{equation}
where $g_{\mu}(x,y)$ is a normalized square integrable wavefunction defining the Hermite-Gaussian wavefunction of order $\mu$; $\mathscr{L}_{z}$ is the quantization length, which disappears after integration, and $(\vec{\epsilon}_{kn})_{i}$ is the $i^{th}$ component of a unit vector whose direction is defined along the polarization of light indexed by momentum $k$ and polarization index $s$. From this, we see that $\mathcal{E}_{(i,n)}(\vec{r})$ gives the spatial dependence of the $\omega=\omega_{n}$ frequency component of the electromagnetic field. In addition, we express the electromagnetic hamiltonian in terms of the electric displacement field $\vec{D}$ in order to make the evolution of the quantum fields fully consistent with Maxwell's equations \cite{quesada2017you}.

Overall, the piezoelectric interaction Hamiltonian is given by the portion of the electromagnetic Hamiltonian that is proportional to strain:
\begin{align}
\hat{H}_{int}&=-\frac{1}{2}h_{ijk}\int d^{3}r \hat{D}_{i}\frac{\partial}{\partial r_{k}}\hat{u}_{j}\\
&=\sum_{mn}\!\hbar g_{ijk(mn)}\!\!\left(\!\hat{c}_{(\!i,m\!)}(\!t\!)\!+\!\hat{c}_{(\!i,m\!)}^{\dagger}(\!t\!)\!\!\right)\!\!\left(\!\hat{b}_{(\!j,n\!)}(\!t\!)\!+\!\hat{b}_{(\!j,n\!)}^{\dagger}(\!t\!)\!\!\right)
\end{align}
Here, $g_{ijk(mn)}$ is the piezoelectric interaction coupling constant between electromagnetic mode $m$ and acoustic mode $n$:
\begin{equation}
\boxed{g_{ijk(mn)}\!\equiv\! i\frac{\sqrt{\!\frac{\omega_{m}}{\omega_{n}}}}{4V_{\!eff}^{(mn)}}\!\!\sqrt{\!\frac{h_{ijk}^{2}}{\eta_{\text{eff}}^{(m)}\rho_{\text{eff}}}}\!\!\int\!\!\! d^{3}r\mathcal{E}_{(i,m)}\!(\vec{r})\frac{\partial w_{(j,n)}\!(\vec{r})}{\partial r_{k}}}
\end{equation}
and we have consolidated electromagnetic and acoustic mode volumes into an electromechanical mode volume: $V_{\text{eff}}^{(mn)}=\sqrt{V_{\text{eff}}^{EM(m)}V_{\text{eff}}^{mech(n)}}$.

Comparing this to our original hamiltonian for the total system and its simplified equations of motion, we have:
\begin{equation}
\sqrt{\gamma_{ex}}\approx -\frac{2 i g_{EM}\sqrt{\Gamma_{ex}}}{\Gamma}
\end{equation}
Where $g_{EM}=i g_{ijk(mn)}$, we obtain a real value for $\sqrt{\gamma_{ex}}$. In general, $\gamma_{ex}$ is a measured quantity rather than derived, but we now see how changing the piezoelectric coupling $g_{EM}$ affects it.

\subsection{Reciprocal relationship between photoelasticity and electrostriction}\label{appc}
From the thermodynamics of dielectric media, the differential change in total internal energy $U$ of a bulk dielectric is given by:
\begin{equation}
dU = TdS + X_{ij}dx_{ij} + E_{i}dD_{i}
\end{equation}
which can be Legendre-transformed to the Helmholtz Free energy $A\equiv U-TS$ (leaving $D$ and $x$ independent):
\begin{equation}
dA = -SdT + X_{ij}dx_{ij} + E_{i}dD_{i}
\end{equation}
which yields the Maxwell relation:
\begin{equation}\label{maxwellRelation}
\left(\frac{\partial X_{ij}}{dD_{k}}\right)_{x,T,\bar{D}}=\left(\frac{\partial E_{k}}{dx_{ij}}\right)_{T,D,\bar{x}}
\end{equation}
While this kind of relation was used previously to describe the connection between the forward and converse piezoelectric effects, it can also be used in non-piezoelectric materials to describe the connection between photoelasticity and electrostriction. Expressing $E$ in terms of the inverse permittivity $\eta$:
\begin{equation}
E_{k}=\eta_{k\ell}D_{\ell},
\end{equation}
and expanding the derivative of $E$ with respect to strain in the Maxwell relation \eqref{maxwellRelation}  using the product rule, we have:
\begin{equation}
\left(\frac{\partial X_{ij}}{dD_{k}}\right)_{x,T,\bar{D}}=\left(\frac{\partial \eta_{k\ell}}{dx_{ij}}\right)_{T,D,\bar{x}}\!\!\!\!\!\!\!\!\!\!\!\! D_{\ell} \;\;-\;\;\eta_{k\ell}\left(\frac{\partial D_{\ell}}{dx_{ij}}\right)_{T,D,\bar{x}}
\end{equation}
The derivative in the first term on the right hand side is the photoelastic coefficient $p_{k\ell ij}$, and the second term on the right-hand side vanishes in a non-piezoelectric material. From this, we obtain the following relation for the photoelastic coefficient:
\begin{equation}
\left(\frac{\partial X_{ij}}{dD_{k}}\right)_{x,T,\bar{D}}=\frac{p_{k\ell ij}}{\epsilon_{0}} D_{\ell}
\end{equation}
If we assume the photoelasticity tensor $p_{ijk\ell}$ to be a constant of the material, we can differentiate this expression again to have the form for $p_{k\ell ij}$:
\begin{equation}
\frac{p_{k\ell ij}}{\epsilon_{0}}=\left(\frac{\partial^{2}X_{ij}}{\partial D_{\ell}\partial D_{k}}\right)=\left(\frac{\partial^{3}F}{\partial D_{\ell}\partial D_{k}\partial x_{ij}}\right)=\left(\frac{\partial^{3}F}{\partial x_{ij}\partial D_{\ell}\partial D_{k}}\right)=\left(\frac{\partial^{2}E_{k}}{\partial x_{ij}\partial D_{\ell}}\right)=\left(\frac{\partial \eta_{k\ell}}{\partial x_{ij}}\right)\equiv \frac{p_{k\ell ij}}{\epsilon_{0}}
\end{equation}
This shows that the converse to the photoelastic effect is the generation of stresses due to the application of an applied electric field (which is the only source of the electric displacement field in this system). Unlike the piezoelectric effect, this  stress is proportional to the \emph{square} of the electric field, and is known as electrostriction \cite{osterberg1937piezodielectric,boyd2008nonlinear}.

\section{Acoustic vs Optical Phonons: How optical phonons contribute negligibly to overall strain}\label{appb}
In this Appendix, we use \cite{kittel1986introduction} and \cite{cline2017variational} to show how at wavelengths much longer than the crystal lattice unit cell (such as how micron-scale sound waves at GHz-scale frequencies in solids are much longer than the typically nm-scale unit cells), optical phonons do not contribute significantly to strain in the material. We accomplish this by showing that in this regime, all optical phonon modes are ones where the center of mass of the unit cell is approximately stationary, which implies the lattice spacing between unit cells is also constant, and that the material undergoes approximately no strain.

To understand this, we refer to the one-dimensional case of a coupled lattice of harmonic oscillators similar to the nearest-neighbor example discussed in chapter 12 of \cite{cline2017variational} and to the discussion of lattice vibrations in chapter 4 of \cite{kittel1986introduction}. In the limit of small perturbation where these inter-atomic couplings are approximately harmonic, the full 3D hamiltonian can be decomposed as the sum of three independent 1D hamiltonians, which lends generality to this treatment. In short, the dynamics for a system of masses $\{m_{i}\}$ within unit cells indexed by $j$, coupled by potential energy $U$ is given by the following lagrangian:
\begin{equation}
\mathscr{L}= T-U=\sum_{ij\ell}\frac{1}{2}T_{j\ell}^{(i)} \dot{q}_{j}^{(i)}\dot{q}_{\ell}^{(i)} - \sum_{ij\ell}\frac{1}{2}\Big(V_{j\ell}^{(i)}q_{j}^{(i)}q_{\ell}^{(i)}+V_{j\ell}^{(i,i-1)}q_{j}^{(i)}q_{\ell}^{(i-1)}+V_{j\ell}^{(i,i+1)}q_{j}^{(i)}q_{\ell}^{(i+1)}\Big)
\end{equation}
Here, $q_{j}^{(i)}$ is the coordinate of the $j^{th}$ atom in the $i^{th}$ unit cell of the crystal. The kinetic energy tensor $T_{j\ell}^{(i)}$ is given by:
\begin{equation}
T_{j\ell}^{(i)}=\delta_{j\ell} m_{j}^{(i)}
\end{equation}
where $m_{j}^{(i)}$ is the mass of the $j^{th}$ atom in the $i^{th}$ unit cell. The potential energy tensors $V_{j\ell}^{(i)}$ and $V_{j\ell}^{(i,i\pm1)}$ are given by:
\begin{equation}
V_{j\ell}^{(i,i+\epsilon)}=\bigg(\frac{\partial^{2}U}{\partial q_{j}^{(i)}\partial q_{\ell}^{(i+\epsilon)}}\bigg|_{q=0}
\end{equation}
where $\epsilon$ may be either $0$ to describe the portion of an atom's potential energy due to interactions within its own unit cell, or $\pm 1$ to describe the portion of the atom's potential energy due to interactions with neighboring unit cells.

From this lagrangian, we have the equations of motion:
\begin{equation}
\frac{\partial U}{\partial q_{j}^{(i)}}=-\frac{d}{dt}\Big(\frac{\partial T}{\partial \dot{q}_{j}^{(i)}}\Big)
\end{equation}
which with the kinetic and potential energy tensors become:
\begin{equation}
\sum_{\ell} (V_{j\ell}^{(i)}q_{\ell}^{(i)}+V_{j\ell}^{i,i+1}q_{\ell}^{i+1} + V_{j\ell}^{(i,i-1)}q_{\ell}^{i-1})+\sum_{\ell}T_{j\ell}^{(i)}\ddot{q}_{j}^{(i)}=0
\end{equation}

Using $a$ as the distance between corresponding atoms in adjacent unit cells, and the trial propagating wave solution:
\begin{equation}
q_{n}^{(N)}(t)=u_{n}^{(N)}e^{ikNa}e^{-i\omega t}
\end{equation}
we can can decouple the equations of motion from adjacent unit cells using $q_{n}^{(N\pm 1)}=q_{n}^{(N)}e^{\pm ika}$ to obtain the linear relation:
\begin{equation}
\sum_{\ell}(V_{j\ell}- \omega^{2}T_{j\ell})u_{\ell}^{(N)}=0
\end{equation}
where now
\begin{equation}
V_{j \ell}= V_{j\ell}^{(i)} + V_{j\ell}^{(i,i-1)}e^{-ika} + V_{j\ell}^{(i,i+1)}e^{ika}
\end{equation}
It is straightforward to understand that $V_{j\ell}^{(i)}$ is Hermitian (due to equality of mixed partial derivatives and a real-valued potential energy). Moreover, the total $V_{j\ell}$ including the terms containing $V_{j\ell}^{(i,i-1)}$ and $V_{j\ell}^{(i,i+1)}$ is also hermitian since the sum of these two terms taken together, is equal to its own conjugate transpose. Based on the hermiticity of $V$, we can derive an orthogonality relation for the normal modes at different frequencies (see \cite{cline2017variational}):
\begin{equation}
(\omega_{r}^{2}-\omega_{s}^{2})u_{j[r]}^{(N)*}T_{j\ell}u_{\ell[s]}^{(N)}=0
\end{equation}
where $u_{j[r]}^{(N)}$ is the $j^{th}$ component of the normal mode at frequency $\omega_{r}$.

To prove that optical phonons do not contribute significant strain, we start in the long wavelength limit of $k=0$. In this limit, we can prove that $\omega=0$ is an eigenfrequency of this system, for which the eigenvector $\vec{u}_{(\omega=0)}$ at $k=0$ is given by $(1,1,1,...,1)/\sqrt{n}$. By multiplying $V_{j\ell}$ by a differential displacement $du_{\ell}^{(N)}$ instead of $u_{\ell}^{(N)}$, this is equivalent to the zero-frequency-zero-momentum eigenmode being one where all of the forces acting on each atom are fully canceled out (adding to zero). This normal mode is an acoustic mode where all atoms move together. 

From the orthogonality relation above, we then find that at $k=0$, all other normal modes (which have frequencies other than zero, and generally quite large) must obey the condition
\begin{equation}
\sum_{j} m_{j}u_{j}^{(N)}=0
\end{equation}
or in short, that the center of mass of the unit cell is stationary. Because the center of mass of each unit cell is stationary at these nonzero frequencies, the spacing between unit cells is also stationary, so that there are no oscillations in compression, tension, shear, or any other form of stain in the material. In short, these optical phonon modes contribute no overall strain to the material. Here we distinguish the overall strain as being viewed at macroscopic length scales that average over many unit cells.

When $k$ is nonzero, but still small enough that the wavelength is much larger than the unit cell, the acoustic mode is still one where the components are nearly equal in sign and magnitude. As a consequence, the optical modes will still be nearly stationary, and will not contribute significantly to the overall strain of the material. It is because of this that we need only consider acoustic phonons in the optomechanical interaction. The driving laser may excite optical phonons as well (i.e., Raman scattering), but this is a loss mechanism not related to the transduction process itself.

\textcolor{black}{\section{Quick note on simplification of strain tensor as single derivative}\label{appStraingradient}
In general, the strain tensor $x_{ij}$ is mathematically defined as \cite{landaulifshitz2021}:
\begin{equation}
    x_{ij}\equiv \frac{1}{2}\left(\left(\frac{du_{i}}{dr_{j}}+\frac{du_{j}}{dr_{i}}\right) +\frac{du_{k}}{dr_{i}}\frac{du_{k}}{dr_{j}}\right).
\end{equation}
Conceptually, the strain tensor applies to deformable extended bodies whose coordinates are given by a range of values for $\vec{r}$. When under strain, points of the body formerly at location $\vec{r}$ are displaced to locations $\vec{r}'$ which will vary as a function of $\vec{r}$. The displacement function $\vec{u}(\vec{r})=\vec{r}'-\vec{r}$ defines this distortion, and we may express the differential change of a line segment of length $d\ell$ within this body to $d\ell'$ via the relation:
\begin{equation}
d\ell'^{2} = d\ell^{2} + 2x_{ij}dr_{i}dr_{j}.
\end{equation}
This implicitly defines $x_{ij}$ as the strain tensor shown above.}
\textcolor{black}{For small enough strains that these first derivatives are much less than unity, we may neglect products of these derivatives to obtain the \emph{linearized} strain tensor:
\begin{equation}
x_{ij}= \frac{1}{2}\left(\frac{\partial u_{i}}{\partial r_{j}} + \frac{\partial u_{j}}{\partial r_{i}}\right).
\end{equation}
We are free to define the antisymmetric counterpart tensor $\bar{x}_{ij}$ to $x_{ij}$ such that:
\begin{equation}
\bar{x}_{ij}= \frac{1}{2}\left(\frac{\partial u_{i}}{\partial r_{j}} - \frac{\partial u_{j}}{\partial r_{i}}\right)= \frac{1}{2}\left((\vec{\nabla}\times \vec{u})_{k\neq i,j}\right),
\end{equation}
which as shown corresponds to components of the curl of the displacement. Summing together both tensors gives us the simple relation:
\begin{equation}
\frac{\partial u_{i}}{\partial r_{j}}=x_{ij} + \bar{x}_{ij}.
\end{equation}}

\textcolor{black}{From Stokes' Theorem \cite{Boas2006MathMethods}, we can write the curl in $\bar{x}_{ij}$ as the limit of a line integral of the displacement dotted along a closed path
\begin{align}
\int_{A}(\vec{\nabla}\times\vec{u})\cdot d\vec{a} &= \oint_{\partial A}\vec{u}\cdot d\vec{r},\\
(\vec{\nabla}\times\vec{u})\cdot \hat{n} &= \lim_{A\rightarrow 0}\frac{1}{A}\oint_{\partial A}\vec{u}\cdot d\vec{r},
\end{align}
where the surface of area $A$ has boundary $\partial A$; $d\vec{a}$ is a vector given by unit vector $\vec{n}$ perpendicular to  the surface $A$ multiplied by the magnitude differential area $da$; and $d\vec{r}$ is a differential vector pointing along the curve defining the boundary $\partial A$ of $A$.}

\textcolor{black}{From the preceding equation, we see that if there is no rigid body rotation in the system, then the line integral of the displacement $\vec{u}$ around any closed path in the  system must be zero, and therefore that $(\vec{\nabla}\times\vec{u})=0$.}

\textcolor{black}{Finally, where $(\vec{\nabla}\times\vec{u})=0$, we have that $\bar{x}_{ij}=0$ and therefore, that:
\begin{equation}\label{rigidApprox}
    \frac{\partial u_{i}}{\partial r_{j}}=x_{ij}.
\end{equation}
Note that the assumption of no rigid body rotation leading to \eqref{rigidApprox} also applies on the differential scale. In short, this approximation also excludes any incidence of torsion waves in the system, where parallel sheets of the system may be twisting with respect to each other in an oscillatory fashion. One may imagine a transduction mechanism using a significant optomechanical coupling between these torsion waves in a multimode optical waveguide, and orbital angular momentum modes of light, but this is beyond the scope of this work.}

\end{widetext}
\end{document}